\documentclass[preprint]{elsarticle}

\usepackage{graphicx}
\usepackage{hyperref}
\usepackage{latexsym}
\usepackage[english]{babel}
\usepackage[squaren]{SIunits}

\usepackage{amssymb,amsfonts,amsmath,amsbsy,theorem,amscd}
\usepackage{textcomp}
\usepackage{color}
\usepackage{xcolor,colortbl}
\usepackage{todonotes}
\usepackage{enumerate}

\def\bsymbol#1{\mbox{\boldmath$\displaystyle#1$\unboldmath}}

\newcommand{\bfnu}{{\textbf{n}}}

\numberwithin{equation}{section}

\newcommand{\eps}{\ensuremath{\varepsilon}}

\newcommand{\Div}{\operatorname{div}}

\newcommand{\habil}[1]{}

\newcommand\bu{{\mathbf v}}
\newcommand\bs{{ s}}

\newtheorem{claim*}{Claim}

\theorembodyfont{\rmfamily}

\newenvironment{proof*}[1]{{\bf Proof #1:}}{\hspace*{\fill}\rule{1.2ex}{1.2ex}\\ }

\newcommand{\bfzeta}{{\bsymbol{\zeta}}}

\usepackage[percent]{overpic}
\usepackage{subfigure}
\usepackage{tikz}
\usepackage{booktabs}
\usepackage{url}
\usepackage[font=normalsize,labelfont=bf]{caption}
\usepackage{textcomp}

\begin{document}

\begin{frontmatter}

\title{A Phase-Field Model for Fluid-Structure Interaction}

\author[label1]{Dominic Mokbel}
\ead{dominic.mokbel@tu-dresden.de}
\author[label2]{Helmut Abels}
\ead{helmut.abels@mathematik.uni-regensburg.de}
\author[label1,label3]{Sebastian Aland}
\ead{sebastian.aland@htw-dresden.de}
\address[label1]{Insitut f\"ur wissenschaftliches Rechnen, Technische Universit\"at Dresden, 01062 Dresden, Germany}
\address[label2]{Fakult\"at f\"ur Mathematik, Universit\"at Regensburg, 93040 Regensburg, Germany}
\address[label3]{Fakult\"at Informatik/Mathematik, Hochschule f\"{u}r Technik und Wirtschaft Dresden, 01069 Dresden, Germany}

\begin{keyword}
fluid-structure interaction, phase-field, diffuse interface, viscoelasticity, contact problem, fully eulerian
\end{keyword}

\begin{abstract}
In this paper, we develop a novel phase-field model for fluid-structure interaction (FSI), that is capable to handle very large deformations as well as topology changes like contact of the solid to the domain boundary. 
The model is based on a fully Eulerian description of the velocity field in both, the fluid and the elastic domain. Viscous and elastic stresses in the Navier-Stokes equations are restricted to the corresponding domains by multiplication with their characteristic functions. To obtain the elastic stress, an additional Oldroyd-B - like equation is solved. 
Thermodynamically consistent forces are derived by energy variation. The convergence of the derived equations to the traditional sharp interface formulation of fluid-structure interaction is shown by matched asymptotic analysis.
The model is evaluated in a challenging benchmark scenario of an elastic body traversing a fluid channel. A comparison to reference values from Arbitrary Lagrangian Eulerian (ALE) simulations shows very good agreement. 
We highlight some distinct advantages of the new model, like the avoidance of re-triangulations and the stable inclusion of surface tension. Further, we demonstrate how simple it is to include contact dynamics into the model, by simulating a ball bouncing off a wall. We extend this scenario to include adhesion of the ball, which to our knowledge, cannot be simulated with any other FSI model. 
While we have restricted simulations to fluid-structure interaction, the model is capable to simulate any combination of viscous fluids, visco-elastic fluids and elastic solids. 
\end{abstract}

\end{frontmatter}

\section{Introduction}
In fluid-structure interaction (FSI) problems, a  solid structure interacts with an
internal or surrounding fluid.  Such problems arise in many scientific and engineering applications, 
for example in aeroelasticity, sedimentation, biological fluids and biomechanics, see \cite{Review} for a recent review. 
Yet the modeling of FSI problems is mathematically challenging due to the fundamental differences of the involved materials: a continually deforming (i.e. flowing) fluid, as opposed to a structurally rigid solid whose atoms are tightly bound to each other. 

Most modeling approaches deal with this discrepancy by introducing two different coordinate systems and two numerical meshes for the two materials. The most popular approach is the Arbitrary Lagrangian-Eulerian (ALE) method \cite{Hughes} in which the computational domain is subdivided into a fluid domain $\Omega_f$ and a structure domain $\Omega_s$. 
The elastic structure is described in the Lagrangian coordinate system, with deformations captured in a displacement vector field. On the contrary, the fluid domain is described in an Eulerian coordinate system and the variable of interest is the velocity field. 
Both meshes are aligned at the fluid/solid interface which is typically moved with the calculated velocity. 
While this methodology provides a sound mathematical description and leads to a very accurate domain representation, it also comes with limitations on the evolution of the solid structure, which breaks down for large deformations or large translational and rotational movements. 
Hence, even simple scenarios like an elastic body moving through a viscous fluid may become impossible to simulate, since technically advanced (re-)triangulation methods and clever interpolations are needed. This has led to the development of alternative modeling approaches in recent years, in particular full Eulerian formulations \cite{FreiRichterWick} and interface capturing methods. 

In interface capturing methods, the fluid and solid domain and their respective interface are implicitly described by an additional field variable. Most popular interface capturing methods are the level-set \cite{Sethian}, volume-of-fluid \cite{Hirt}, and phase-field method \cite{Aland}. 
Only in the recent decade numerical schemes have been developed to describe fluid–structure interactions in level-set methods \cite{CottetMaitreMilcent,Legay,Qiao} and volume-of-fluid methods \cite{Li}.

Phase-field models provide an alternative interface capturing approach which may offer, depending on the application of interest,  some distinct advantages. 
An auxiliary field variable $\phi$, the phase-field, is introduced and used to indicate the phases, e.g., $\phi=1$ in the solid and $\phi=-1$ in the fluid. The phase-field function varies smoothly between these distinct values across the interface, resulting in a small but finite interface thickness. Depending on the application of interest, phase-field methods offer some distinct advantages over other interface capturing methods. For example, they can intrinsically include mass conservation and transport stabilization and 
allow for fully discrete energy stable schemes, see e.g. \cite{Chen,Gruen}
, for two-phase flows.
Further these methods offer a simple mechanism to couple the multi-phase system to additional physical processes, for example on the interface or in the bulk phases, see for examples \cite{nihms,Lowengrub,Garcke2014a}. 
In case of a non-negligible surface energy, phase-field methods allows for a monilithic coupling of interface advection and flow equations, which can prevent time step restrictions due to stiff interfacial forces \cite{ifdim}.

While phase-field methods are state-of-the-art for multi-phase flow problems, only one preliminary approach has been made to provide a phase-field model for FSI \cite{Sun2014}. 
This might be due to the fact, that the interface in phase-field method is diffuse with a finite thickness, making it harder to combine the solid and the fluid material in a consistent way.
Here, we present an improved phase-field method for fluid-structure-interaction along with analytical and numerical validation. 

We start by recapitulating the sharp interface equations for FSI in Sec.~\ref{sec:fsi sharp}. The equations are extended to a phase-field formulation in Sec.~\ref{sec:pfmodeling}. The model is based on a thermodynamically consistent derivation (Sec. \ref{sec:EnergyDissipation}). We provide 
a formal sharp interface limit showing convergence of the derived equations to traditional FSI formulations (Sec. \ref{sec:sharp interface asymptotics}). Numerical tests are presented in Sec.~\ref{sec:numericaltests}. Special focus is put on a comparison to the ALE reference solution of an elastic cell traversing a cylindrical channel. 
After this validation, we demonstrate the potential of the method in Sec.~\ref{sec:potential}, by simulating 
\begin{enumerate}[(i)]
\item a solid object moving through a fluidic channel without grid remeshing,
\item FSI with strong surface energy, i.e. surface tension forces,
\item contact dynamics of a bouncing ball, and
\item adhesion of an elastic object to a rigid wall.
\end{enumerate}
The paper closes with conclusions on the applicability of the method in Sec.~\ref{sec:conclusion}.

\section{Phase-Field Model for FSI}

\subsection{Sharp Interface Model}
\label{sec:fsi sharp}
Before deriving the phase-field model, let us begin by presenting the sharp interface equations for FSI. 
Let the computational domain $\Omega\subseteq\mathbb{R}^n$ be divided into a fluid domain and a solid domain. 
To be consistent with the later phase-field model, we call these domains $\Omega_{-1}$(fluid) and $\Omega_1$(solid). 
We introduce a common velocity field ${\bf v}:\Omega\rightarrow\mathbb{R}^n$ to indicate movements of the fluid and the solid material. Further, let us denote the material derivative by $\partial^{\bullet} = \partial_t + {\bf v}\cdot \nabla$.
Throughout this work, we consider incompressible elastic materials, i.e. of $1/2$ Poisson ratio, although the model can be easily adapted to account for compressible solids. 
Balance laws for mass and momentum yield the evolution equations
\begin{align}
\partial^{\bullet}(\rho_i {\bf v}) &=
\nabla \cdot \mathbb{S}_i +\nabla p  & \text{in }\Omega_i \label{navier stokes sharp} \\
\nabla \cdot {\bf v} &= 0 & \text{in }\Omega_i  \label{incompress sharp} 
\end{align}
for $i=-1,1$, where ${\bf v}, p, \mathbb{S}_i$ denote the velocity, pressure and phase- dependent stress. 

To describe the elastic stress in the Eulerian framework we introduce the left Cauchy-Green strain tensor $\sigma$. 
In an undeformed configuration, this strain tensor is equal to the identity tensor $\mathbb{I}$. During movement of the elastic material the strain tensor memorizes deformations by following the evolution equation
\begin{align}
\partial^\bullet \sigma - \nabla{\bf v}^T \cdot \sigma - \sigma\cdot \nabla{\bf v} &= 0 & {\rm in}~ \Omega_1 \label{B evolution solid sharp}
\end{align}
The left-hand side of Eq.~\eqref{B evolution solid sharp} is also known as the upper-convected Maxwell time derivative that rotates and stretches with the deformation. 
The corresponding elastic stress is $\mu_1(\sigma-\mathbb{I})$, where $\mu_1$ is the shear modulus of the elastic material. 
In a fluid, the elastic stress vanishes since there is no strain, i.e. 
\begin{align}
\sigma&=\mathbb{I} & {\rm in}~ \Omega_{-1}  \label{sigma equals I}
\end{align}
for all times.
The total phase-dependent stress is then given by the elastic stress plus a viscous part,
\begin{align}
\mathbb{S}_i &= \nu_i(\nabla{\bf v}+\nabla{\bf v}^T) + \mu_i(\sigma-\mathbb{I}) \label{stress sharp}
\end{align}
where $\nu_i$ is the viscosity  and $\mu_i$ is the shear modulus of the respective phase. 
In particular $\mu_{-1}=0$ in a purely viscous fluid, and $\nu_1=0$ if the elastic solid has no additional viscosity. 

Finally two jump conditions are specified at the fluid/solid interface,
\begin{align}
[{\bf v}]_-^+ = 0, \qquad -[\mathbb{S}]_-^+\cdot {\bf n} + [p]_-^+{\bf n}  = \gamma\kappa {\bf n}, \label{jump cond}
\end{align}
where $[f]_-^+=f_{+1}-f_{-1}$ denotes the jump in $f$ across the interface. 
 The first is the continuity of the velocity. The second condition is the interfacial force balance including a possible surface tension force at the fluid/solid interface with surface tension $\gamma$ and total curvature $\kappa$.

\subsection{Phase-field modeling}
\label{sec:pfmodeling}
Let $\phi$ denote a phase-field that distinguishes between the fluid domain ($\phi\approx -1$) and the solid domain ($\phi\approx 1$) within a computational domain $\Omega$. Hence $\Omega_i$ is approximated by the domain where $\phi\approx i$.
Since the phase-field approach allows mixing of the two domains to a certain degree, we define the velocity field now to be the volume-averaged velocity of this mixture, see \cite{AGG} for details. 
The density is chosen as a linear combination of the densities in the two phases: $\rho(\phi) = \rho_1(1+\phi)/2  + \rho_{-1}(1-\phi)/2 $.
Balance laws for mass and momentum yield the evolution equations for phase-field and velocity:
\begin{align}
\partial^{\bullet} \phi &= -\nabla\cdot {\bf J} & \text{in }\Omega \label{transport}  \\
\partial^{\bullet}(\rho(\phi) {\bf v}) &=
\nabla \cdot \mathbb{S} (\phi) +\nabla p + {\bf F}  & \text{in }\Omega \label{navier stokes} \\
\nabla \cdot {\bf v} &= 0 & \text{in }\Omega  \label{incompress} 
\end{align}
where the stress depends now on the phase-field and the force ${\bf F}$ and flux ${\bf J}$ are specified later to meet the requirement of non- increasing energy. 

To obtain an equation for the diffuse elastic strain tensor, we need to combine Eqs.~\eqref{sigma equals I} and \eqref{B evolution solid sharp}.
A typical approach in phase-field modeling is to multiply an equation with a characteristic function of its domain (here: $\Omega_{-1}$ or $\Omega_1$) and to extend the domain then to the larger computational domain (here: $\Omega$), see \cite{li2009solving}. We follow a similar approach here and multiply Eq.~\eqref{sigma equals I} with a function $\alpha(\phi)$ and Eq.~\eqref{B evolution solid sharp} with a function $\lambda(\phi)$. Adding up both results, we obtain the common equation for the diffuse elastic strain tensor as
\begin{align}
\lambda (\phi) \left(\partial^\bullet \sigma - \nabla{\bf v}^T \cdot \sigma - \sigma\cdot \nabla{\bf v} \right)  + 
\alpha(\phi) (\sigma-\mathbb{I}) = 0. \label{B evolution combined}
\end{align}
For $\alpha=0$ the equation reduces to the strain evolution for an elastic solid \eqref{B evolution solid sharp}, while for $\lambda=0$ it reduces to the strain description of a fluid, i.e., Eq.~\eqref{sigma equals I}. 
In case $\alpha=1$, Eq.~\eqref{B evolution combined} is also known as Oldroyd-B equation. This equation is used to describe Maxwell-type visco-elasticity with $\lambda$ being the relaxation time controlling the dissipation of elastic stress. 
The above generalization of the Oldroyd-B equation to arbitrary $\alpha$, leads to the relaxation time $\lambda/\alpha$. Note, that this ratio is the only free parameter of Eq.~\eqref{B evolution combined}, but the introduction of $\alpha$ effectively allows to choose this ratio equal to infinity by setting $\alpha=0$. 

The total phase-dependent stress is then given by the elastic stress plus a viscous part,
\begin{align}
\mathbb{S}(\phi) &= \nu(\phi)(\nabla{\bf v}+\nabla{\bf v}^T) + \mu(\phi)(\sigma-\mathbb{I}) \label{stress}
\end{align}
where $\nu(\phi)$ is the viscosity  and $\mu(\phi)$ is the shear modulus of the respective phase. 
We use linear interpolations for all phase-dependent quantities, 
\begin{align*} 
\mu(\phi) &= \mu_1(1+\phi)/2  + \mu_{-1}(1-\phi)/2 \\
\nu(\phi) &= \nu_1(1+\phi)/2  + \nu_{-1}(1-\phi)/2 \\
\lambda(\phi) &= \lambda_1(1+\phi)/2  + \lambda_{-1}(1-\phi)/2 \\
\alpha(\phi) &= \alpha_1(1+\phi)/2  + \alpha_{-1}(1-\phi)/2
\end{align*}
Hence to model interaction of a solid ($\phi=1$) with a fluid ($\phi=-1$), we insert the given physical parameters for elastic shear modulus $\mu_1$, the fluid viscosity $\nu_{-1}$ and set $\mu_{-1}=\nu_1=\alpha_1=\lambda_{-1}=0$.
It remains to specify $\alpha_{-1}$ and $\lambda_1$ which can be interpreted to control the relaxation time in the diffuse interface region where a mixture of fluid and elastic phase is present. Note, that due to the structure of Eq.~\eqref{B evolution combined} the only free parameter here is the ratio $\lambda_1/\alpha_{-1}$, which we can also think of an interface relaxation time. Obviously, this time has to be scaled somehow with the characteristic time scale $T$ of the considered problem, which suggests to simply use $\lambda_1/\alpha_{-1}=T$, for example by setting $\alpha=1, \lambda=T$. This parameter choice will also be numerically tested in Sec. \ref{sec:results}.

An overview of the parameters can be found in Tab. \ref{tab:parameters}. 
Note, that also visco-elastic material phases can be modeled. For Kelvin-Voigt visco-elasticity it suffices to add a viscosity inside of the elastic material. For Maxwell visco-elasticity one can just prescribe the Maxwell relaxation time for $\lambda$. 
Hence, any combination of two phases, be it viscous, visco-elastic or solid, can be modeled by choosing the parameters as given in Tab. \ref{tab:parameters}.

\begin{table}
\begin{center}
\begin{tabular}{cccc|l} \rowcolor[gray]{.8}
viscosity&shear modulus& relax. time& &corresponding material  \\
\rowcolor[gray]{.8} ~~~~$\nu(\phi)$~~~~ &~~~~ $\mu(\phi)~~~~$ & $\lambda(\phi)$ & $\alpha(\phi)$ &\\
\hline	
\rowcolor[gray]{.9}  $*$ & $0$  & $0$ & $1$ &viscous fluid \\
\rowcolor[gray]{.9}  $0$  & $*$ & $T$  &$0$ & elastic solid   \\ 
\rowcolor[gray]{.9}  $*$ & $*$ & $T$  &$0$ & visco-elastic Kelvin-Voigt	 \\ 
\rowcolor[gray]{.9}  $0$  & $*$ & $*$ &$1$ & visco-elastic Maxwell   
\end{tabular}
\end{center}
\caption{Different material laws can be obtained by different choice of parameters. The star symbol $'*'$ marks parameters that are given by the physical problem itself, $T$ indicates the characteristic time scale of the given problem.}
\label{tab:parameters}
\end{table}

\subsection{Energy Dissipation}
\label{sec:EnergyDissipation}
To close the system of equations, it remains to specify the flux ${\bf J}$ and force ${\bf F}$ to obtain a thermodynamically consistent evolution. We define the total energy $E$ of the system as sum of kinetic, elastic \cite{Boyaval} 
and diffuse surface energy \cite{Aland_SPP_Abschluss_2017} as follows,
\begin{align}
E=&\int_{\Omega}{\frac{\rho(\phi)}{2}\left| \mathbf{v}\right|^2} 
+{\frac{\mu(\phi)}{2}\text{tr}\left(\sigma-\ln{\sigma}-\mathbb{I}\right)}
+\tilde\gamma\left(\frac{\epsilon}{2}\left|\nabla\phi\right|^2+\frac{1}{\epsilon}W(\phi)\right) ~\text{d}x. \label{energy}
\end{align}  
Here, '$\text{tr}(A)$' is the trace of a Matrix $A$, $\tilde\gamma = 3/2\sqrt{2}\cdot\gamma$ the (scaled) surface tension, $\epsilon$ the interface thickness and $W(\phi)=\frac{1}{4}(1-\phi^2)^2$ a double-well potential. 
Inserting \eqref{navier stokes}-\eqref{stress} into Eq.~\eqref{energy} we can compute the time evolution of the energy. The complete computation is carried out in the appendix. We obtain
\begin{align}
\frac{d}{dt} E = &\int_{\Omega} -\frac{\nu(\phi)}{2}\left|\nabla{\bf v}+\nabla{\bf v}^T\right|  \nonumber \\
&\qquad +{\bf J}\cdot\nabla\left[\frac{\mu'(\phi)}{2}\text{tr}(\sigma-\ln\sigma-\mathbb{I}) +\tilde{\gamma}\left(\frac{1}{\epsilon}W'(\phi) - \epsilon\Delta\phi\right) \right] \nonumber \\
& \qquad +{\bf v}\cdot\left[{\bf F} + \nabla\cdot(\rho'(\phi){\bf v}\otimes{\bf J}) + \epsilon\tilde\gamma\nabla\cdot(\nabla\phi\otimes\nabla\phi)\right]   ~\text{d}x. \nonumber \\
&- \int_{\Omega\backslash\{\lambda=0\}}  \frac{\mu(\phi)\alpha(\phi)}{2\lambda(\phi)} \text{tr}(\sigma+\sigma^{-1}-2\mathbb{I}) ~\text{d}x \label{dt energy 1}
\end{align}
under appropriate boundary conditions. Note, that the last term is non-negative for any given tensor $\sigma$ and bounded as explained in Sec.~\ref{sec: energy time derivative}. Hence, with the choice 
\begin{align}
{\bf F} &= -\nabla\cdot(\rho'(\phi){\bf v}\otimes{\bf J}) - \epsilon\tilde\gamma\nabla\cdot(\nabla\phi\otimes\nabla\phi), \label{F}\\
{\bf J} &= -m(\phi) \nabla\left[\frac{\mu'(\phi)}{2}\text{tr}(\sigma-\ln\sigma-\mathbb{I}) + \tilde\gamma\left(\frac{1}{\epsilon}W'(\phi) - \epsilon\Delta\phi\right)  \right] \label{J}
\end{align}
for some mobility function $m(\phi)> 0$, we obtain non-increasing energy,
\begin{align*}
\frac{d}{dt} E &= -\int_{\Omega} \frac{\nu(\phi)}{2}\left|\nabla{\bf v}+\nabla{\bf v}^T\right| +\frac{1}{m(\phi )}|{\bf J}|^2  ~\text{d}x \\
&~~~ -  \int_{\Omega\backslash\{\lambda=0\}}  \frac{\mu(\phi)\alpha(\phi)}{2\lambda(\phi)} \text{tr}(\sigma+\sigma^{-1}-2\mathbb{I}) ~\text{d}x \\
&\leq~ 0.
\end{align*}

\subsection{Governing equations}

In the following we summarize the governing equations for the thermodynamically consistent visco-elastic phase-field model. 
As noted earlier, the model can be used to describe any combination of viscous, visco-elastic and elastic materials, by choosing the parameters accordingly, see. Tab.~\ref{tab:parameters}.

\begin{align}
\label{eq:mom} 
 \partial^\bullet(\rho(\phi) {\bf v}) + \frac{\rho_{-1}-\rho_1}{2}\nabla\cdot\left({\bf v}\otimes m(\phi)\nabla q\right) \\ -\nabla\cdot\left(\nu(\phi)\left(\nabla{\bf v} + \nabla{\bf v}^T\right)+\mu(\phi)\left(\sigma-\mathbb{I}\right)\right)-\nabla p &= -\tilde\gamma\epsilon\nabla\cdot\left(\nabla
\phi \otimes \nabla \phi\right),\nonumber\\
\nabla\cdot{\bf v}&=0,\label{eq:div0}\\
\partial^\bullet\phi&=\nabla\cdot\left(m(\phi)\nabla q\right),\label{eq:transport}\\
\frac{\mu'(\phi)}{2}\text{tr}(\sigma-\ln\sigma-\mathbb{I}) + \tilde\gamma\left(\frac{1}{\epsilon}W'(\phi) - \epsilon\Delta\phi\right)&=q,\label{eq:q}\\
\lambda (\phi) \left(\partial^\bullet \sigma - \nabla{\bf v}^T \cdot \sigma - \sigma\cdot \nabla{\bf v} \right)  + 
\alpha(\phi) (\sigma-\mathbb{I}) &= 0, \label{eq:OldB}
\end{align}

In numerical tests we find, that if the chemical potential $q$ as defined in Eq.~\eqref{eq:q} is used in the evolution equation of the phase-field, the resulting $\phi$ does not provide a good description of the interface layer because of the contributions of the elastic strain. Since the primary purpose of $\phi$ is to track the two-phase interface, we use a simplified version of the $q$ in the numerical simulations, which omits the strain-dependent terms,
\begin{align}
q = \frac{1}{\epsilon} W'(\phi) - \epsilon\Delta \phi \label{eq:q2}
\end{align}
replacing Eq.~\eqref{eq:q}.
 This amounts into a classical advected Cahn-Hilliard equation for $\phi$. 
Note that the resulting system is no longer variational and does not necessarily decrease the energy. However, 
this effect tends to be higher order since away from the interface 
$W'(\phi)=\Delta\phi=0$
and near the interface $\phi$ locally equilibrates yielding 
$W'(\phi) \approx \epsilon^2\Delta\phi$ and thus $q\approx 0$. Note that if $q=0$, then the energy is non-increasing, $\frac{d}{dt} E \leq 0$ which follows from Eq.~\eqref{dt energy 1}.

In the following Section we derive relations of our phase-field model in the sharp interface limit (i.e. $\epsilon\rightarrow 0$) with the aid of formal asymptotic expansions. We perform the analysis for the full model containing the strain dependent terms in $q$, but we will see that the asymptotic analysis holds equally for the simplified model involving Eq.~\eqref{eq:q2}. 

\section{Sharp Interface Asymptotics}
\label{sec:sharp interface asymptotics}

In the following we only consider the case $\lambda (-1)=0,\alpha(-1)=0, \lambda(1)=T>0, \alpha(1)=0$, which corresponds to the coupling of a viscous fluid and an elastic solid.
Following \cite{AGG}, we perform formally matched asymptotic expansions.  
Therefore we consider a solution $\left({\bf v}, p, \phi, q, \sigma\right)$ of the system given by Eqs. \eqref{eq:mom}-\eqref{eq:OldB}.
For the mobility $m$ we distinguish two cases in the following: 
$$
m(\phi) = \begin{cases}
\epsilon m_0 &\mbox{case I}\,,\\
 m_1(1-\phi^2)_+ &\mbox{case II}
\end{cases}
$$
where $m_0,m_1 >0$ are constants and $(.)_+$ is the positive part of the quantity in the
brackets. A lot of calculations and arguments follow closely \cite{AGG} with suitable modifications. For the convenience of the reader we include them in detail although some are the same as in \cite{AGG}.  \\

\subsection{Outer expansions}
The first step is to expand the solution in regions away from the interface. Therefore we assume an expansion of the form
\begin{align}
 {\bf v}^\epsilon =
\sum^\infty_{k=0} \epsilon^k{\bf v}_k, ~ \phi^\epsilon =
\sum^\infty_{k=0} \epsilon^k\phi_k,~\dots.
\end{align}
An expansion of Eq. \eqref{eq:q2} at order $\frac{1}{\epsilon}$ leads to $W'\left(\phi_0\right)=0$, which yields the stable solutions $\phi_0=\pm 1$. 
Expanding the Eq. \eqref{eq:mom} we obtain
\begin{align}
\partial^{\bullet}\left(\rho_i{\bf v}_0\right)-\nabla\cdot\left(\nu_i\left(\nabla{\bf v}_0+\nabla{\bf v}_0^T\right)+\mu_i\sigma_0\right)+\nabla p_0&=0 ~~~\text{ in } \Omega_{i}, \\
\nabla\cdot {\bf v}_0&=0 ~~~\text{ in } \Omega_{i},\\
\label{outerRegion}
\sigma_0&=\mathbb{I}~~~\text{ in }\Omega_{-1},\\
\label{innerRegion}
\partial^\bullet \sigma_0 - \nabla{\bf v}_0^T \cdot \sigma_0 - \sigma_0\cdot \nabla{\bf v}_0&=0~~~\text{ in }\Omega_1,
\end{align}
where $i=-1,1$ and $\Omega_i$ the domain where $\phi_0=i$. Furthermore we denoted by $\rho_{\pm 1}=\rho\left(\pm 1\right)$, $\nu_{\pm 1}=\nu\left(\pm 1\right)$ and $\mu_{\pm 1}=\mu\left(\pm 1\right)$.
Note that we recover the sharp interface equations given in Sec.~\ref{sec:fsi sharp}. 
\subsection{Inner expansions}
As a second step we prepare the expansion in the interface region and therefore introduce new coordinates in a neighborhood of the smoothly evolving interface $\Gamma=\Gamma\left( t\right),~t\geq 0$. We define a local parameterization of $\Gamma$ by
\begin{align}
\bfzeta :I\times U\rightarrow\mathbb{R}^n
\end{align}
with a time interval $I\subset\mathbb{R}$ and a spatial parameter domain $U\subset\mathbb{R}^{n-1}$. The unit normal to $\Gamma\left( t\right)$ will be denoted by $\bfnu$ and points into $\Omega_1$. In the following we adopt the notation from \cite{AGG}. We consider the signed distance function $d\left(t,x\right)$ of a point $x$ to the sharp interface $\Gamma^0\left( t\right)$ with $d\left( t,x\right)>0$ if $x\in\Omega_1\left( t\right)$. In addition, we denote by $z=\frac{d}{\epsilon}$ a rescaled distance. We now introduce the new coordinates by defining a local parameterization of $I\times \mathbb{R}^n$ close to $\bfzeta\left(I\times U\right)$ as follows:
\begin{align}
G^{\epsilon}\left(t,s,z\right)=\left(t,\bfzeta\left( t,s\right)+\epsilon z{\bf \bfnu}\left(t,s\right)\right)
\end{align}
with $s\in U$. It will turn out that we need the following identities, containing a scalar function $b\left(t,x\right)$ and a vector field ${\bf j}\left(t,x\right)$:
\begin{align}
\frac{d}{dt} b (t,x)&=
 -\tfrac{1}{\epsilon}\mathcal{V} \partial_z\hat{b}+\,\,\mbox{h.o.t.}\, ,\\
 \label{eq:innerGrad}
  \nabla_x b &= \nabla_{\Gamma_{\epsilon z}} \hat{b}
 +\tfrac{1}{\epsilon} \partial_z \hat{b} \,\bfnu ,\\
  \nabla_x \cdot\mathbf{j} &=
 \Div_{\Gamma_{\epsilon z}}\hat{\mathbf{j}}
 +\tfrac{1}{\epsilon} \partial_z \hat{\mathbf{j}} \cdot\bfnu,\\
  \Delta_x b &= \Delta_{\Gamma_{\epsilon z}}\hat{b}-\tfrac{1}{\epsilon}(\kappa+\epsilon
 z|\mathcal{S}|^2)\partial_z\hat{b}+\tfrac{1}{\epsilon^2}\partial_{zz}\hat{b}+\,\,\mbox{h.o.t.}\,,
\end{align}
with the correspondences
\begin{itemize}
\item $\hat{b}$ is the denotation of $b$ in the new coordinates with\\$\hat{b}\left(t,s\left(t,x\right),z\left(t,x\right)\right)=b\left(t,x\right)$. 
\item $\mathcal{V}=\partial_t\bfzeta\cdot\bfnu$ is the scalar normal velocity.
\item h.o.t. stands for higher order terms as $\eps\to 0$.
\item $\nabla_x$ is the gradient with respect to the spatial variables.
\item $\nabla_{\Gamma_{\epsilon z}}$ is the surface gradient on
 $\Gamma_{\epsilon z}:=\{\bfzeta(\bs)+\epsilon z\bfnu(\bs)\mid \bs\in
 U\}$.
 \item $\Div_{\Gamma_{\epsilon z}}\hat{\mathbf{j}}$ is the
 divergence of $\hat{\mathbf{j}}$ on $\Gamma_{\epsilon z}$.
 \item $\kappa$ is the mean curvature of $\Gamma(t)$.
 \item $|\mathcal{S}|$ is the spectral norm of the Weingarten map $\mathcal{S}$ of $\Gamma(t)$,
\end{itemize}
cf.~\cite{AGG}.
Note, that we omit the time dependence in the following argumentations, as done for $\Gamma_{\epsilon z}$ above. 
Moreover, we will make use of the relations (see Appendix of \cite{AGG})
 \begin{align*}
 \nabla_{\Gamma_{\epsilon z}}\hat{b} (\bs,z) &= \nabla_\Gamma
 \hat{b}(\bs,z)+\,\,\mbox{h.o.t.}\,,\\
 \Div_{\Gamma_{\epsilon z}}  
 \hat{\bf{j}}(\bs,z)&=
 \Div_{\Gamma}
 \hat{{\bf j}}(\bs,z)+\,\,\mbox{h.o.t.}\,,\\
 \Delta_{\Gamma_{\epsilon z}}
   \hat{b}(\bs,z)&=\Delta_\Gamma\hat{b}(\bs,z)+\,\,\mbox{h.o.t.}\, ,
 \end{align*} where
$\nabla_\Gamma,
 \Div_{\Gamma},\Delta_\Gamma$ are the respective surface operators on $\Gamma$. 
 
 \subsection{Matching conditions}
 As for the outer variables, we now assume an $\epsilon$-series approximation for the unknown functions $\left({\bf V},P,\Phi,Q,\Sigma\right) $ in the inner variables:
 \begin{align*}
	  {\bf V}^\epsilon =
\sum^\infty_{k=0} \epsilon^k{\bf V}_k, ~ \Phi^\epsilon =
\sum^\infty_{k=0} \epsilon^k\Phi_k,~\dots ~.
 \end{align*}  
Representatively, we obtain the following matching conditions for the phase field function at $x=\bfzeta\left(s\right)$:
\begin{align}\label{match1}
\underset{z\to\pm\infty}{\lim} \Phi_0(z,s)&=\phi_0(x\pm)\,,\\
\label{match2}
\underset{z\to\pm\infty}{\lim} \partial_z\Phi_1(z,s)
&=\nabla\phi_0(x\pm)\cdot\bfnu~,
\end{align} 
where $\phi_0(x\pm)$ denotes the limit
$\underset{\delta\searrow 0}{\lim}\,\phi_0(x\pm\delta\bfnu)$. 

\subsection{The equations to leading order}

We insert the asymptotic expansions into Eqs.\
\eqref{eq:mom}-\eqref{eq:OldB} and ask that each individual coefficient
of a power in $\epsilon$ vanishes. The leading order of equation \eqref{eq:q}
is $\frac{1}{\epsilon}$, which gives
\begin{equation}\label{Leadphi}
0=\partial_{zz}\Phi_0- W'(\Phi_0).
\end{equation}
Using (\ref{match1}) we obtain
\begin{equation}\label{Leadphib}
\Phi_0(z)\to\pm 1\quad\mbox{for}\quad z\to\pm\infty\,.
\end{equation}
We now assume additionally that
\begin{equation*}
\Phi_0(0)=0\,.
\end{equation*}
 Together with this condition (\ref{Leadphi}),
(\ref{Leadphib}) has a unique solution.
Hence $\Phi_0$ does not depend on $t$ and
$s$. 
The leading order of Equation (\ref{eq:div0}) yields
\begin{equation}\label{VLO}
\partial_z{\bf V}_0\cdot\bfnu = \partial_z({\bf V}_0\cdot\bfnu)=0\,.
\end{equation}
The matching condition implies that $({\bf V}_0\cdot\bfnu)(z)$ is
bounded. Hence
\begin{eqnarray*}
(\bu_0\cdot\bfnu)(x+)&=&\underset{z\to\infty}{\lim}({\bf V}_0\cdot \bfnu)(z)
= \underset{z\to-\infty}{\lim}
({\bf V}_0\cdot\bfnu)(z)=(\bu_0\cdot\bfnu)(x-)\,.
\end{eqnarray*}
Thus
\begin{equation*}
[\bu_0\cdot\bfnu]^+_-=0\,,
\end{equation*}
where $[u]^+_-(x)=u(x+)-u(x-)$ denotes the jump of a quantity at the
interface. \\
For the analysis of Eq.\ \eqref{eq:transport} we have to distinguish the different case for the mobility.\\[1ex]
\emph{Case I: $m(\phi)=\epsilon m_0\,.$}\\
Equating the order $\frac{1}{\epsilon}$ term we obtain from Eq.\ \eqref{eq:transport}
\begin{equation}\label{aa33}
-\mathcal{V}
\partial_z\Phi_0+(\bu_0\cdot\bfnu)\partial_z\Phi_0=\partial_z(m_0\partial_zQ_0\bfnu)\cdot\bfnu=m_0\partial_{zz}Q_0\,.
\end{equation}
Moreover, matching yields
\begin{equation*}
\partial_zQ_0\to 0\quad\mbox{for}\quad z\to\pm\infty\,.
\end{equation*}
If we integrate (\ref{aa33}) with respect to $z$, we obtain
\begin{equation*}
\mathcal{V} = \bu_0\cdot\bfnu\,.
\end{equation*}
Since $\partial_{zz}Q_0=0$, we conclude that $Q_0$ is independent of  $z$. \\[1ex]
\emph{Case II: $m(\phi)= m_1(1-\phi^2)_+\,.$}\\
Equating the order $\frac{1}{\epsilon^2}$ terms, we obtain
\begin{equation*}
0=\partial_z(m_1(1-\Phi^2_0)\partial_zQ_0\bfnu)\cdot\bfnu=\partial_z(m_1(1-\Phi^2_0)\partial_zQ_0)\,.
\end{equation*}
Moreover, matching yields
\begin{equation*}
m_1(1-\Phi^2_0)\partial_zQ_0\to 0\quad\mbox{for}\quad z\to\pm\infty\,
\end{equation*}
and therefore
\begin{equation*}
m_1(1-\Phi^2_0)\partial_zQ_0\equiv 0,
\end{equation*}
which implies
\begin{equation*}
Q_0=Q_0(s,t)\,.
\end{equation*}
At order $\frac{1}{\epsilon}$ we obtain
\begin{equation*}
-\mathcal{V}\partial_z\Phi_0+({\bf
  V}_0)\cdot\bfnu\partial_z\Phi_0=0.
\end{equation*}
As before integration yields $\mathcal{V} = \bu_0\cdot\bfnu$. \\

Next we discuss the equation for the conservation of linear momentum.
Application of Eq. \eqref{eq:innerGrad} for each component yields 
\begin{eqnarray*}
\nabla_x\bu&=&\tfrac{1}{\epsilon}\partial_z{\bf V}\otimes\bfnu+
\nabla_{\Gamma_{\epsilon z}}{\bf V}\,,\\
D_x\bu &=& \tfrac{1}{2} \tfrac{1}{\epsilon}
(\partial_z{\bf V}\otimes\bfnu+\bfnu\otimes\partial_z{\bf
  V})+\tfrac{1}{2}(\nabla_{\Gamma_{\epsilon z}}{\bf
  V}+(\nabla_{\Gamma_{\epsilon z}}{\bf V})^\top)\,.
\end{eqnarray*}
For the following we define $\mathcal{E}({\bf A})=\tfrac{1}{2}({\bf A}+{\bf A}^\top)$ for a
quadratic matrix $\bf A$. Hence
\begin{eqnarray*}
\nabla_x\cdot(\nu(\phi)D_x \bu) &=&
\tfrac{1}{\epsilon^2}\partial_z(\nu(\Phi)\mathcal{E}
(\partial_z{\bf V}\otimes\bfnu))\bfnu
+\tfrac{1}{\epsilon}\partial_z(\nu(\Phi)\mathcal{E}(\nabla_{\Gamma_{\epsilon
    z}}{\bf V}))\bfnu\\ &&
+\tfrac{1}{\epsilon}\nabla_{\Gamma_{\epsilon
    z}}\cdot(\nu(\Phi)\mathcal{E}(\partial_z{\bf V}\otimes\bfnu))
+\nabla_{\Gamma_{\epsilon
    z}}\cdot(\nu(\Phi)\mathcal{E}(\nabla_{\Gamma_{\epsilon z}}{\bf
  V}))\\
&=&\tfrac{1}{\epsilon^2}\partial_z(\nu(\Phi)\mathcal{E}(\partial_z{\bf
  V}\otimes\bfnu)\bfnu)
+\tfrac{1}{\epsilon}\partial_z(\nu(\Phi)\mathcal{E}(\nabla_{\Gamma_{\epsilon
    z}}{\bf V})\bfnu)\\
&&+\tfrac{1}{\epsilon}\nabla_{\Gamma_{\epsilon
    z}}\cdot(\nu(\Phi)\mathcal{E}(\partial_z{\bf V}\otimes\bfnu))
+\nabla_{\Gamma_{\epsilon
    z}}\cdot(\nu(\Phi)\mathcal{E}(\nabla_{\Gamma_{\epsilon z}}{\bf
  V}))\,,
\end{eqnarray*}
where we used $\partial_z\bfnu=0$ as in \cite{AGG}. We conclude from Eq. \eqref{VLO} 
\begin{equation*}
(\bfnu\otimes\partial_z{\bf V}_0)\bfnu=(\partial_z{\bf V}_0\cdot\bfnu)\bfnu=0\,.
\end{equation*}
Since $\Phi_0$ is independent of $t$ and $s$,
Eq. \eqref{eq:innerGrad} implies
\begin{equation*}
\nabla\phi\otimes\nabla\phi=\tfrac{1}{\epsilon^2}(\partial_z\Phi_0)^2(\bfnu\otimes\bfnu)
+\tfrac{2}{\epsilon}\partial_z\Phi_1\partial_z\Phi_0(\bfnu\otimes\bfnu)+\,\,\mbox{h.o.t.}
\end{equation*}
Because of $(\nabla_\Gamma \bfnu) \bfnu =0 $, we conclude
\begin{equation*}
\epsilon\nabla\cdot(\nabla\phi\otimes\nabla\phi)=\tfrac{1}{\epsilon^2}\partial_z(\partial_z\Phi_0)^2\bfnu+\tfrac{1}{\epsilon}(\partial_z\Phi_0)^2(\nabla_\Gamma\cdot\bfnu)\bfnu
+\tfrac{1}{\epsilon}\partial_z(\partial_z\Phi_1\partial_z\Phi_0)\bfnu+\,\,\mbox{h.o.t.}
.
\end{equation*}
Since the leading order of the chemical potential does not depend on $z$,
we the term 
$ \Div ({\bf v}\otimes
 m(\phi)\nabla m)$ gives no contribution to the order $\frac
1{\epsilon^2}$.
Therefore the order $\frac{1}{\epsilon^2}$ terms from the
momentum equation yield
\begin{equation}\label{MEQ-2}
\tilde\gamma\partial_z(\partial_z\Phi_0)^2\bfnu
+\partial_z(\nu(\Phi_0)\partial_z{\bf V}_0)=0\,.
\end{equation}
Multiplication of Eq. \eqref{MEQ-2} with $\bfnu$, taking $\partial_z\bfnu=0$ and
$\partial_z{\bf V}_0\cdot\bfnu=0$ into account yields 
\begin{equation*}
\tilde\gamma\partial_z((\partial_z\Phi_0)^2)=0\,.
\end{equation*}
Therefore (\ref{MEQ-2}) implies
\begin{equation}\label{MEQ-2a}
\partial_z(\nu(\Phi_0)\partial_z{\bf V}_0)=0\,.
\end{equation}
The matching conditions yield that ${\bf V}_0(z)$ is bounded. Thus
Eq.\ \eqref{MEQ-2a} interpreted as an ODE in $z$ has only solutions ${\bf V}_0$
which are constant in $z$. Again matching yields 
\begin{equation}
[\bu_0]^+_-=0\,.
\label{vcont}
\end{equation}
Thus we recover the first part of the sharp interface jump condition in Eq.~\eqref{jump cond}.

\subsection{The momentum balance in the sharp interface limit }

Now we analyze the momentum equation to the next order. The
term $\nabla\cdot(\nu(\phi)D\bu)$ gives to the order
$\frac{1}{\epsilon}$,
\begin{equation*}
\partial_z(\nu(\Phi_0)\mathcal{E}(\partial_z{\bf
  V}_1\otimes\bfnu)\bfnu)+\partial_z(\nu(\Phi_0)\mathcal{E}(\nabla_{\Gamma
    }{\bf V}_0)\bfnu)\,.
\end{equation*}
Because of the matching conditions, we require
$\underset{z\to\pm\infty}{\lim}\partial_z{\bf V}_1(z)=\nabla\bu_0(x\pm)\bfnu$.
Hence
\begin{equation}\label{matchv}
\partial_z{\bf V}_1\otimes\bfnu+\nabla_\Gamma{\bf
  V}_0\to\nabla_x\bu\quad\mbox{for}\quad z\to\pm\infty\,.
\end{equation}
Moreover, the term $\Div 
 ({\bf v}\otimes
 m(\phi)\nabla q)$ gives no contribution to order $\tfrac 1\epsilon$.
Thus
we obtain from the momentum equation at order
$\frac{1}{\epsilon}$:
\begin{eqnarray*}
&-&\partial_z(\rho(\Phi_0){\bf
  V}_0)\mathcal{V}+\partial_z(\rho(\Phi_0)({\bf V}_0\otimes{\bf V}_0))\bfnu\\
&-&2\partial_z(\nu(\Phi_0)\mathcal{E}(\partial_z{\bf
  V}_1\otimes\bfnu)\bfnu)-2\partial_z(\nu(\Phi_0)\mathcal{E}(\nabla_{\Gamma
  }{\bf V}_0)\bfnu)
  \\
&-&\tilde\gamma(\partial_z\Phi_0)^2\kappa\bfnu+\tilde\gamma\partial_z(\partial_z\Phi_1\partial_z\Phi_0)\bfnu+\partial_z (\mu(\Phi_0)\Sigma_0)+\partial_zP_0\bfnu=0\,.
\end{eqnarray*}
Integration with respect to $z$ yields after matching and the use of 
Eq.\ \eqref{matchv}
\begin{alignat*}{1}
&-[\rho_0\bu_0]^+_-\mathcal{V}
+[\rho_0\bu_0]^+_-\bu_0\cdot\bfnu-2[\nu\mathcal{E}(\nabla_x\bu_0)]^+_-\bfnu\\
&-\tilde\gamma\biggl(\int^\infty_{-\infty}(\partial_z\Phi_0)^2\textit{d}
z\biggr)\kappa\bfnu
-[\mu(\phi_0)\sigma_0]^+_-\bfnu+[p_0]^+_-\bfnu=0\,.
\end{alignat*}
Since $\bu_0\cdot\bfnu=\mathcal{V}$, we conclude
\begin{equation}
-2[\nu
D\bu_0]^+_-\bfnu+[p_0]^+_-\bfnu-[\mu(\phi_0)\sigma_0]^+_-\bfnu= \tilde\gamma\kappa\bfnu\, .
\label{intstress1}
\end{equation}
Therefore we recover the second part of the sharp interface jump condition in Eq.~\eqref{jump cond}.

\section{Numerical tests}
\label{sec:numericaltests}

Numerical tests are indispensable to validate numerical models and to assess their accuracy. 
Nowadays, the standard benchmark for fluid-structure-interaction is the FeatFlow benchmark \cite{Turek}
where a channel flow induces oscillations of a thin elastic bar attached to a rigid object.
Obviously, the phase-field model is not well suited to represent such a thin structure, since the corresponding interface thickness would be required to be much thinner than the structure itself, resulting in an extremely fine grid and very high computational costs. 
Consequently, we choose a different benchmark system serving the current purpose of testing the phase-field model in a practically relevant situation. 

We consider the flow of a deforming solid ball through a fluid-filled channel. Firstly, this highlights the ability of interface capturing methods to account for movements of the solid and fluid domains with respect to each other. 
Secondly, the test scenario is based on a physically relevant simulation of biological cells traversing a flow channel. Such simulations have been recently established to enable ultra-fast identification of cell mechanical properties \cite{Mokbel2017}. Therefore cells are approximated as homogeneous incompressible elastic solids surrounded by a cortex with an active surface tension. While we neglect this active tension in the benchmark stage, we will add it to the model later to make use of one distinct advantage of the phase-field modeling: the stabilization of stiffness arising between interface advection and surface forces \cite{ifdim}.


\subsection{Test Setup}
\label{sec:setup}

\begin{figure}
\begin{center}
\begin{tabular}{c}
\includegraphics[angle=0,width=0.6\textwidth]{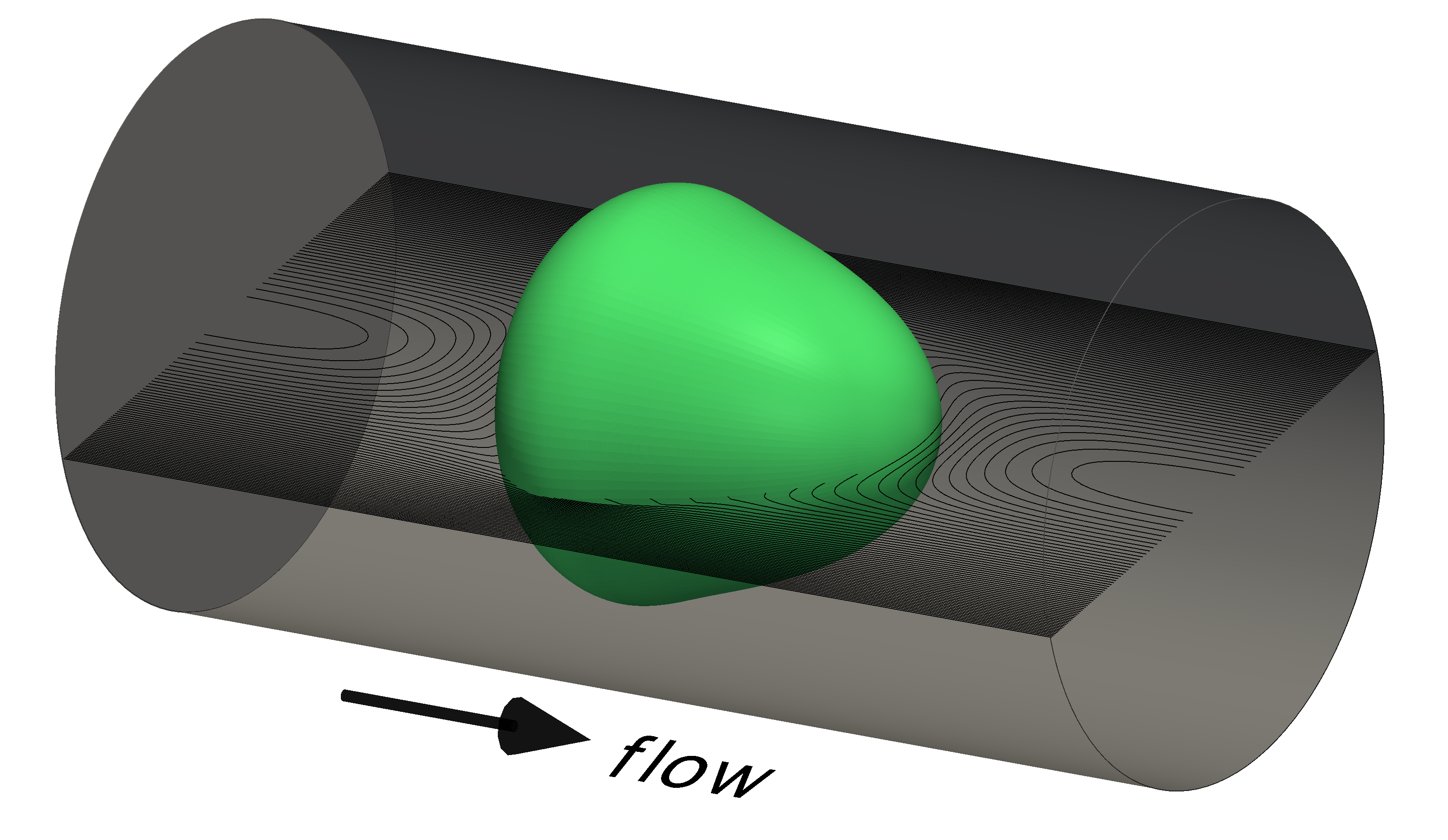}
\end{tabular}
\end{center}
\caption{Illustration of the simulated scenario. An initially spherical elastic cell (green) is deformed by pressure and shear forces as it flows through a fluid-filled cylindrical channel. 
Streamlines (black) visualize fluid movement relative to cell velocity.
}
\label{fig:3D}
\end{figure}

We simulate the flow of an initially spherical solid object, also called {\it cell}, through a fluid-filled channel. 
To be consistent with the reference simulations \cite{Mokbel2017}, we consider the solid object and the channel to be axisymmetric. Axisymmetry effectively reduces the problem to a two-dimensional flow with axisymmetric operators. Hence, we consider a two-dimensional rectangular domain $\Omega$ whose lower boundary represents the symmetry axis, see Fig. \ref{fig:setup} for an illustration.  Here, $\Omega=[0,40]\mu m\times[0,10]\mu m$ which corresponds to a cylindrical channel of radius 10\,\textmu m, see Fig.~\ref{fig:3D} for an illustration.

Periodic boundary conditions are used at channel inlet and outlet for all fields but the pressure, which effectively leads to a channel of infinite length. 
A pressure difference between the channel inlet and outlet is imposed to drive the flow. Consistently with \cite{Mokbel2017} this pressure difference is chosen such that a flow rate of $4\mathrm{e}{-11}\, m^3/s$ appears.
At the channel wall, we impose no slip (${\bf v}=0$) for the velocity and no flux for the Cahn-Hilliard system.

Surface tension forces are neglected at first, but we choose $\tilde\gamma=1$ in Eq.~\eqref{eq:q} such that we keep the stabilizing  Cahn-Hilliard diffusion. 
Moreover, we choose the following physical parameters for our simulations: $\nu_1=\nu_{-1}=10$\,Pa$\cdot$s, $\rho_1=\rho_{-1}=1000\,\frac{\text{kg}}{\text{m}^3}$, $\mu_{-1}=0$. 
The shear modulus of the cell, $\mu_1$ is related to its Young's modulus $E$ by $E=3\mu_1$. Different values of $E$ are used in the tests. The radius of the initially spherical cell is set to cell $r=6\, $\textmu m.

Unless otherwise stated, the standard model parameters are $\epsilon =0.0125\mu\text{m}$ and constant mobility $m=10^{-8}~$m$^3$s/kg. 
The characteristic time scale of the considered problem is approximately 1~ms, we choose the parameters for the Oldroyd-B equation accordingly $\alpha_1 = 0, \alpha_{-1} = 1, \lambda_1 = 1~$ms$, \lambda_{-1} = 0$, as suggested in Tab.~\ref{tab:parameters}. Note, that the only free parameter here is $\lambda_1$, we will later assess the influence of variations in $\lambda_1$ in our numerical tests.

The problem is discretized in the finite element toolbox AMDiS \cite{amdis,Witkowski2015}. 
We chose an adaptive mesh refinement strategy where coarsening and refinement of the mesh are controlled by gradients in the phase-field function. The refinement level at the interface is chosen depending on the choice of $\epsilon$, such that the interface is resolved by at least five degrees of freedom. Away from the interface the larger grid size $h=0.625\mu\text{m}$ is chosen. 

P1 finite elements are used for the pressure and P2 elements for the other variables. 
The time discretization is based on a implicit Euler method. Thereby, the Navier-Stokes and Cahn-Hilliard system are monolithically coupled while the Oldroyd-B equation \eqref{B evolution combined} is solved separately in each time step. 
For details on the axisymmetric equations and the time discretization we refer to the Appendix \ref{sec:timediscrete}. 

\subsection{Benchmark Quantity}
After being deformed by pressure and shear forces, the solid will assume a stationary shape whereupon its flow becomes purely translational. 
We aim in particular to reproduce this state of stationary deformation. Note that this is a highly challenging problem for a phase-field method, since the structure needs to resist any movement, while the fluid keeps flowing around it and continuous movement takes place, even in the diffuse interface.

We introduce the deformation as a measure of the deviation of the cell shape from a circle,
\begin{align*}
d=1-{\rm circularity} = 1- \frac{2\sqrt{A\pi}}{P},
\end{align*}
where $A$ and $P$ denote the area and the perimeter of the 2D view of the deformed object. 
It has been shown in \cite{Mokbel2017} that $d$ is a delicate measure of the cell shape that can be uniquely related to the exact elastic modulus of the cell. We hence use this quantity as a main indicator for comparison of the phase-field model with the ALE reference solution. 

Reference values for the stationary cell shapes are given in \cite{Mokbel2017} for various cell sizes, flow rates and elastic moduli. There, an ALE method was employed using a co-moving grid to keep the cell in the center of the computational domain throughout the simulation. 
These data in \cite{Mokbel2017} has been shown to be extremely accurate in terms of spatial and temporal discretization errors. 
The ALE data has been extensively used to draw comparisons to corresponding experiments. 
The experimental technique, called Real-Time-Deformability Cytometry (RT-DC), can be used to probe mechanical properties of biological cells in flow \cite{Otto}. A validation study with purely elastic spherical particles showed excellent agreement between ALE simulations and experiments. 


\subsection{Simulation Results}
\label{sec:results}

In this section we compare the phase-field method presented in this paper with the ALE reference data and perform a parameter study to justify our choice of $\lambda_1$, $\epsilon$ and $m$, respectively. 

\begin{figure}
\begin{center}
\begin{tabular}{cccc}
\includegraphics[angle=-90,width=0.2\textwidth]{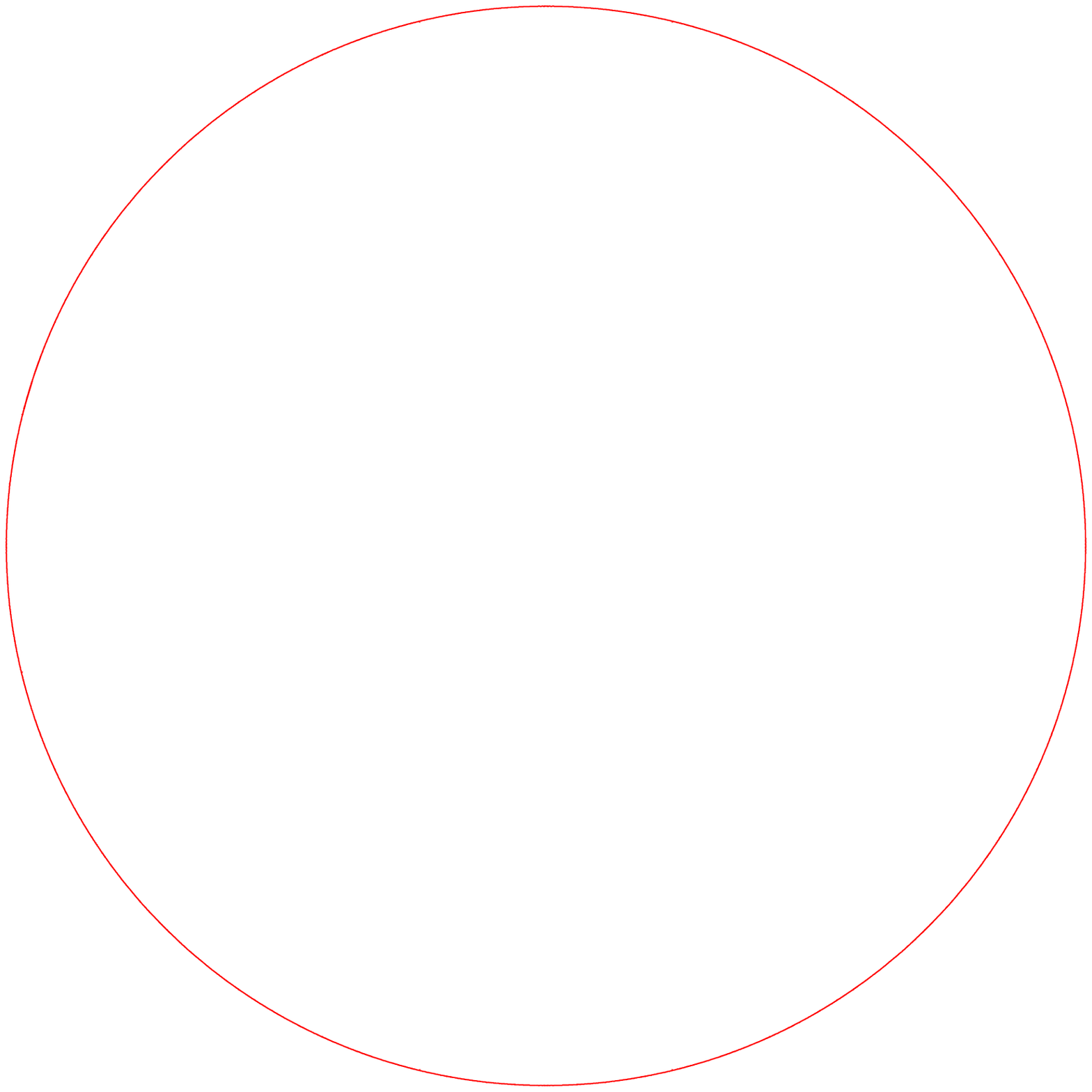} &
\includegraphics[angle=-90,width=0.2\textwidth]{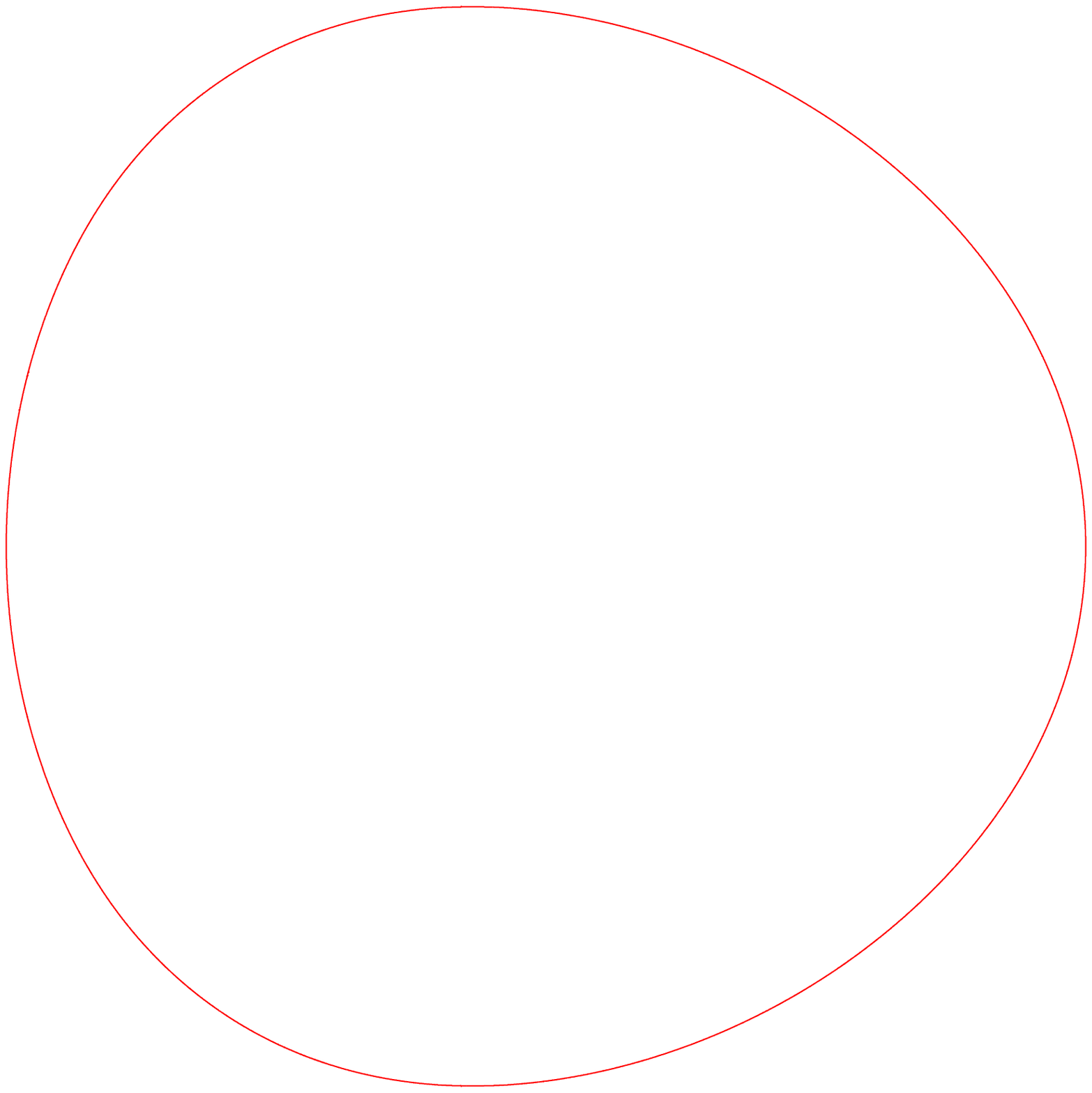} & 
\includegraphics[angle=-90,width=0.2\textwidth]{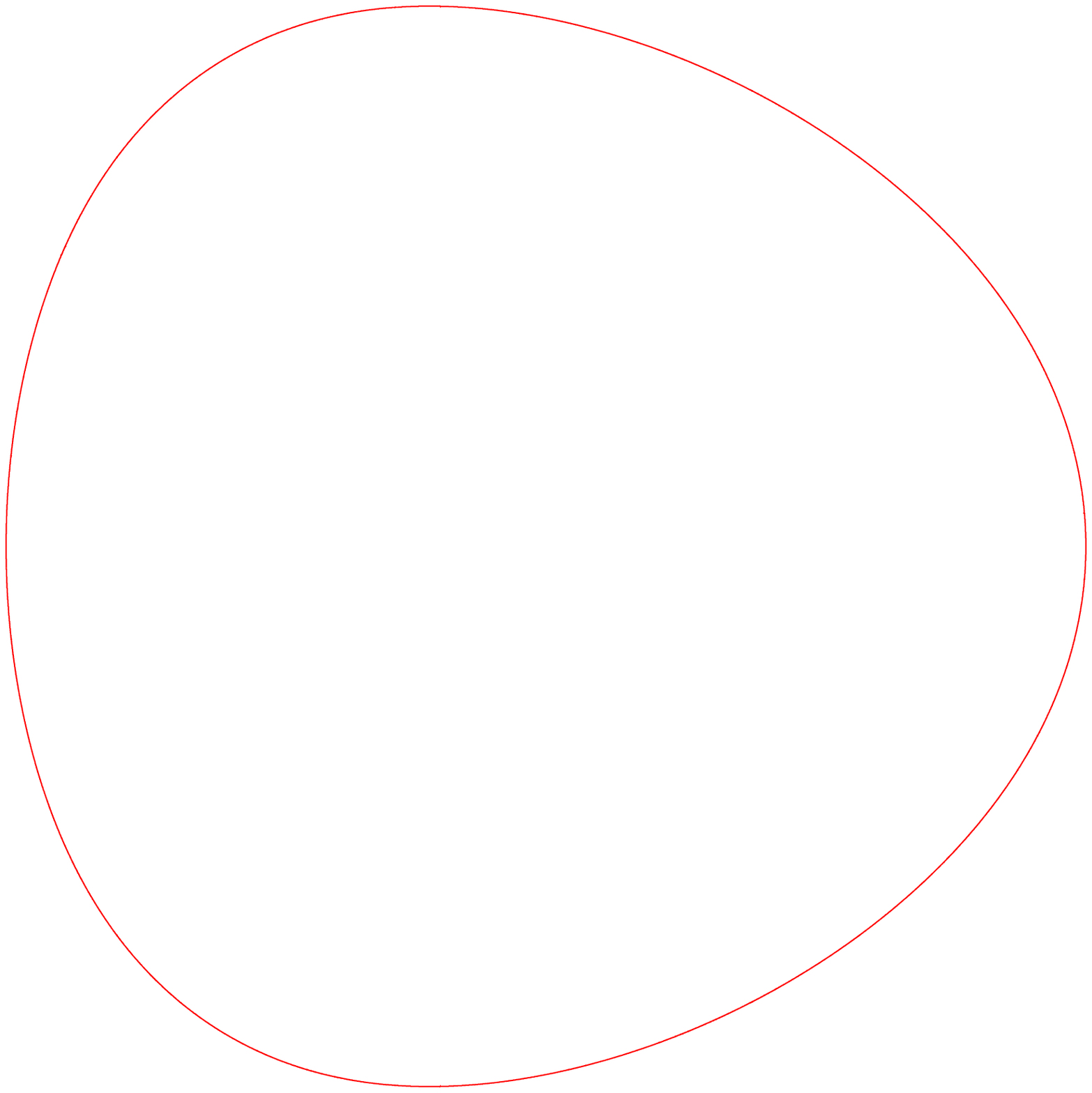} &
\includegraphics[angle=-90,width=0.2\textwidth]{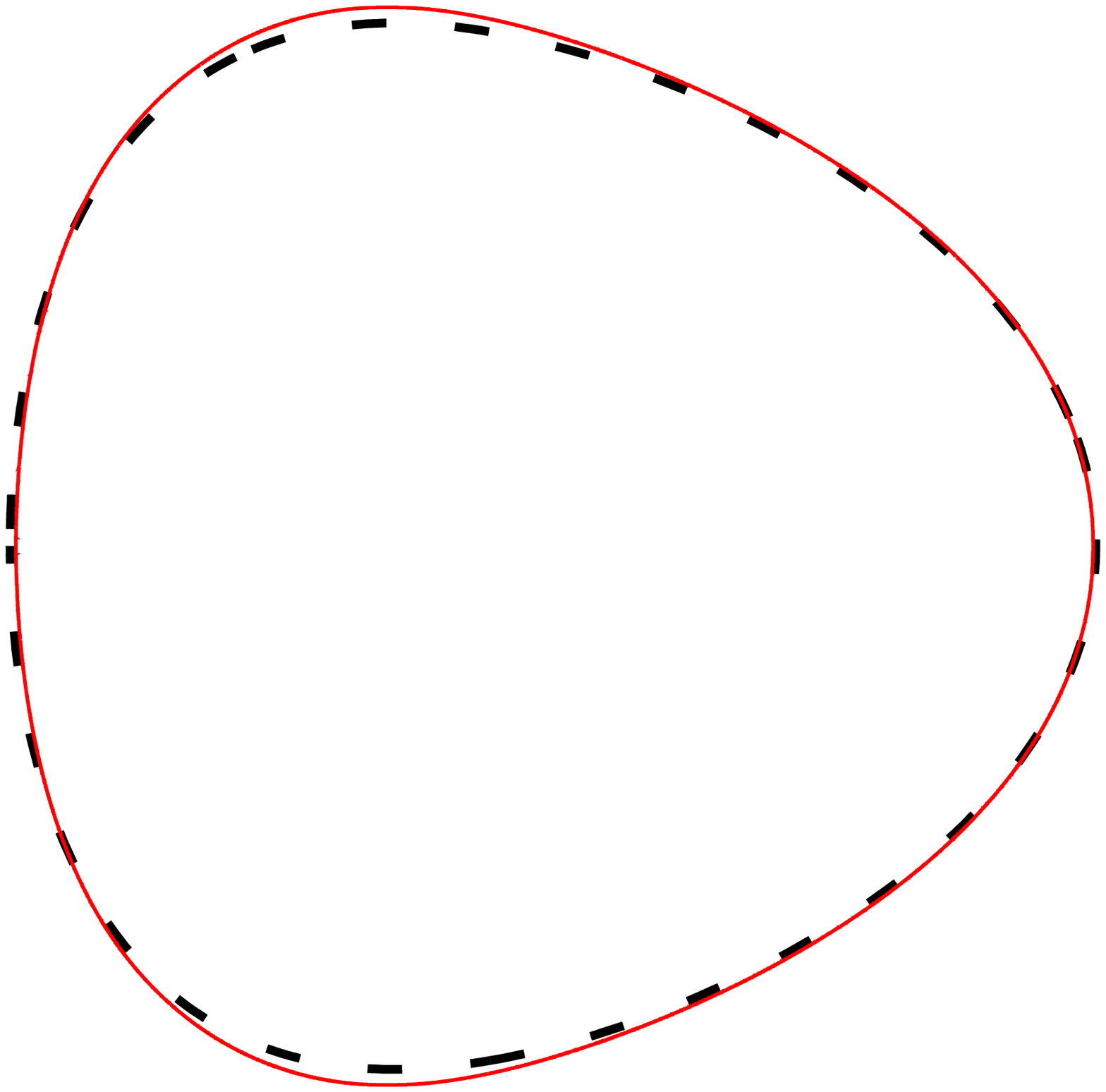} \\
 0\,\textmu s  &  50\,\textmu s & 100\,\textmu s & 3000\,\textmu s
 
\end{tabular}
\end{center}
\caption{Cell shape at different times for $E=6$kPa, $r=6\mu$m, which corresponds to the parameter set discussed in Fig.~\ref{fig:params}. Remarkable changes of the shape occur within the first 100\,\textmu s. The last image compares the stationary shape of the phase-field method with the ALE reference shape (dashed).}
\label{fig:shapes}
\end{figure}

Fig.~\ref{fig:3D} provides an idea of the actually simulated 3D scenario. It illustrates a stationary state shape in the cylinder, together with the streamlines of the flow.
Moreover, Fig.~\ref{fig:shapes} shows the cell shape for different times. It can be seen that the initially spherical cell deforms due to fluid pressure and shear forces until it assumes a quasi-stationary state. 
A comparison with the ALE reference shape shows good agreement (Fig.~\ref{fig:shapes}, right).

\begin{figure}
\begin{center}
\begin{tabular}{cc}
\includegraphics[angle=-90,width=0.48\textwidth]{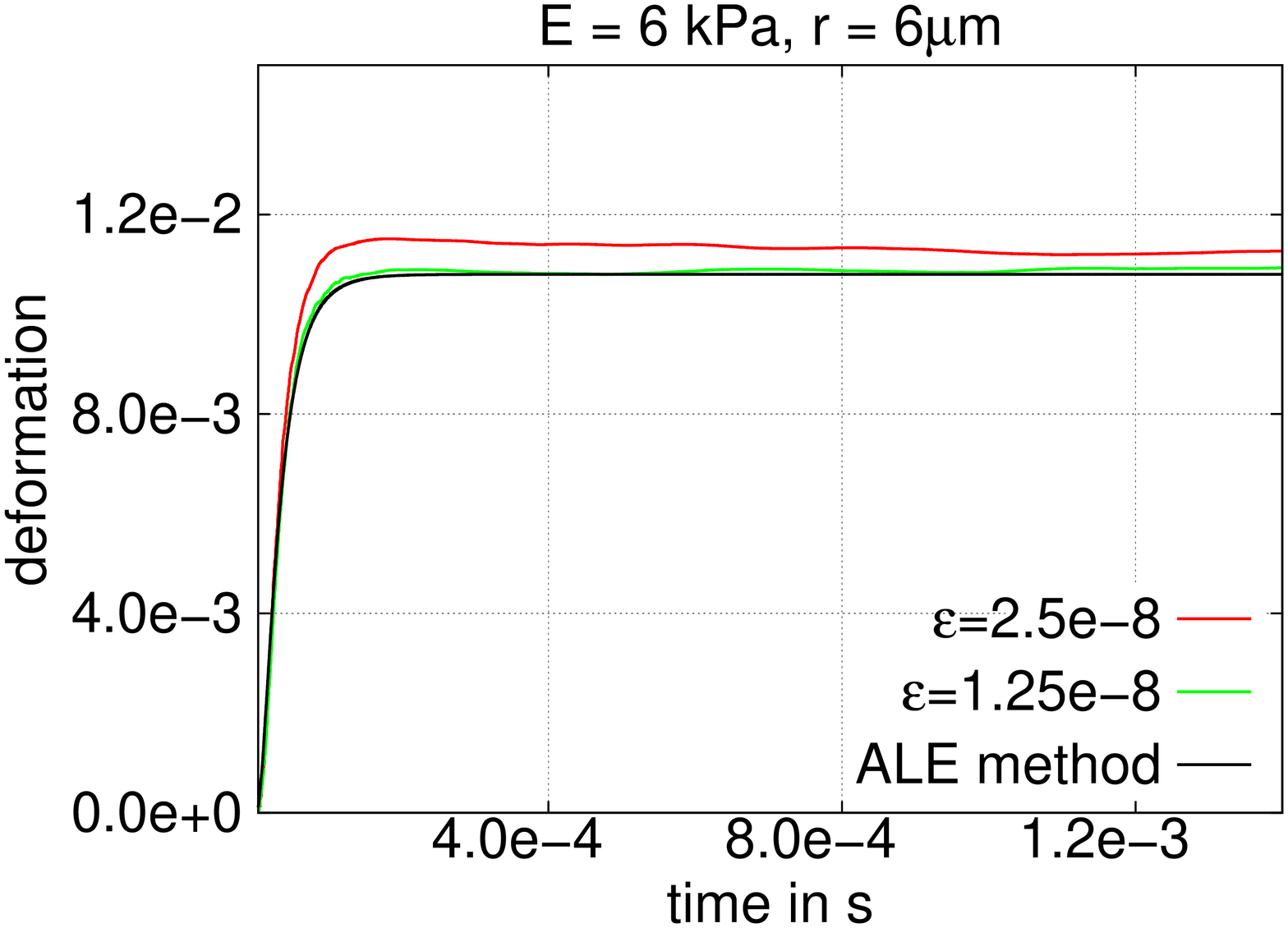} &

\includegraphics[angle=-90,width=0.48\textwidth]{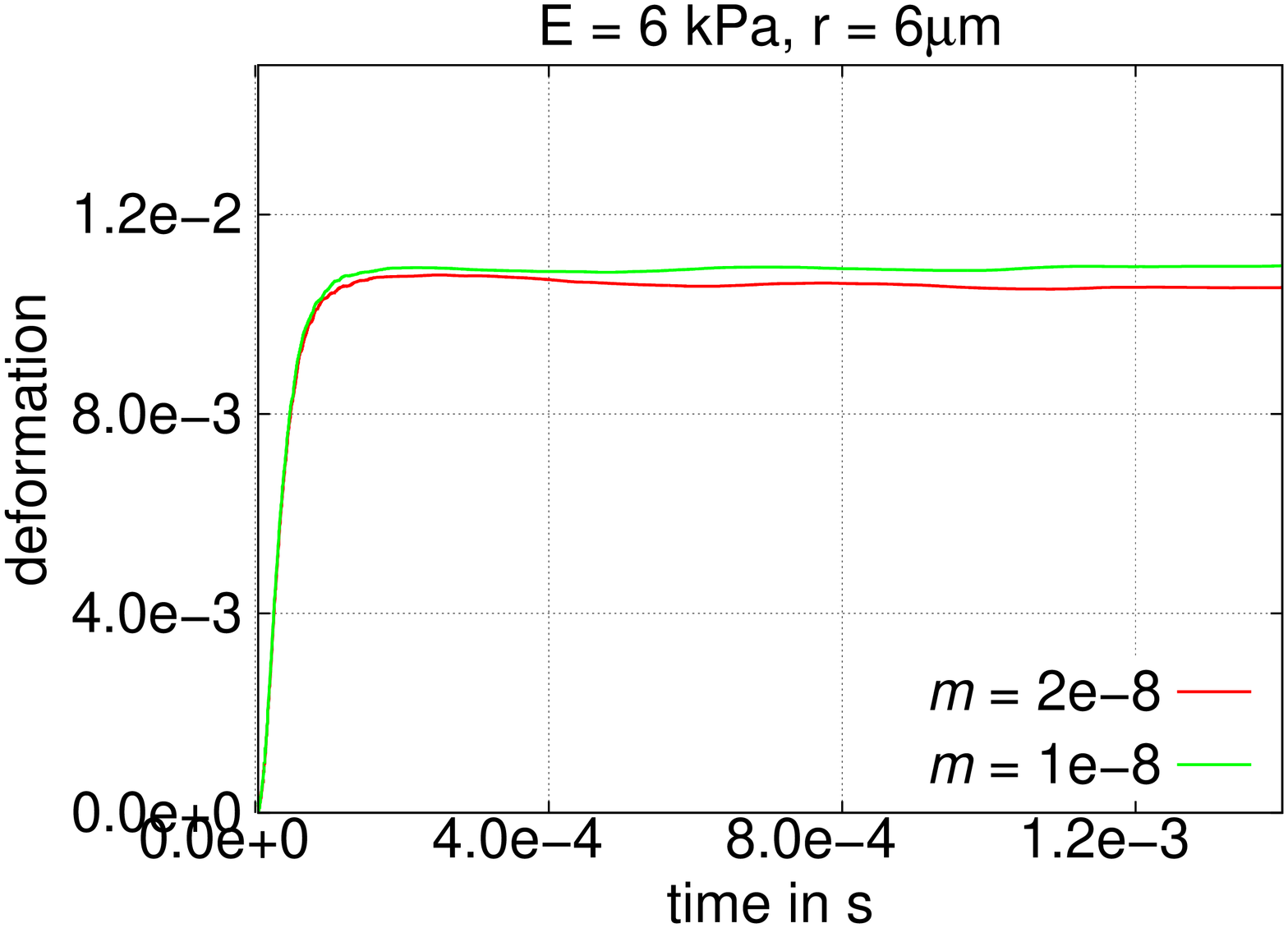} \\
(a)  &  (b)  \\

\includegraphics[angle=-90,width=0.48\textwidth]{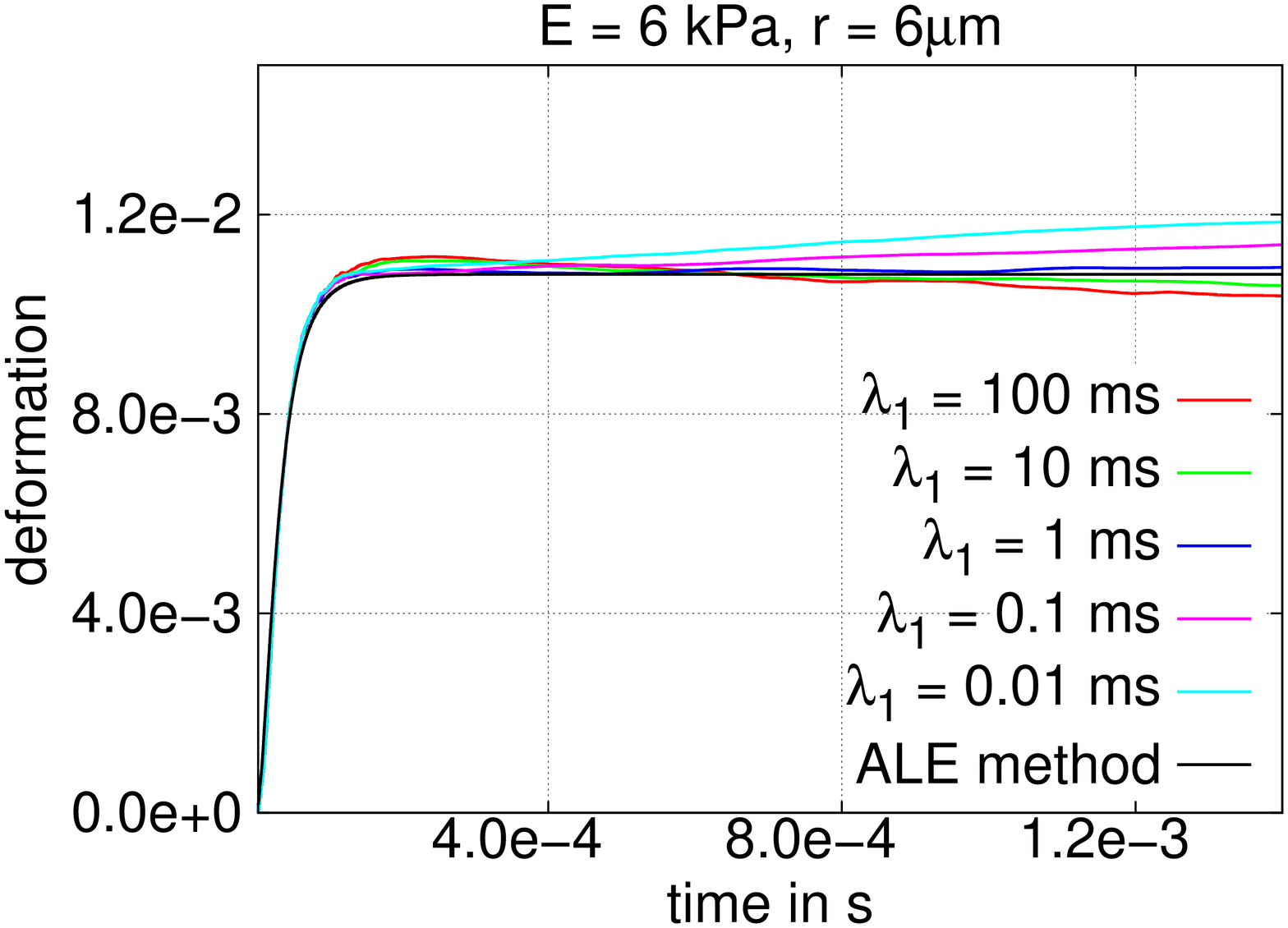} &

\includegraphics[angle=-90,width=0.48\textwidth]{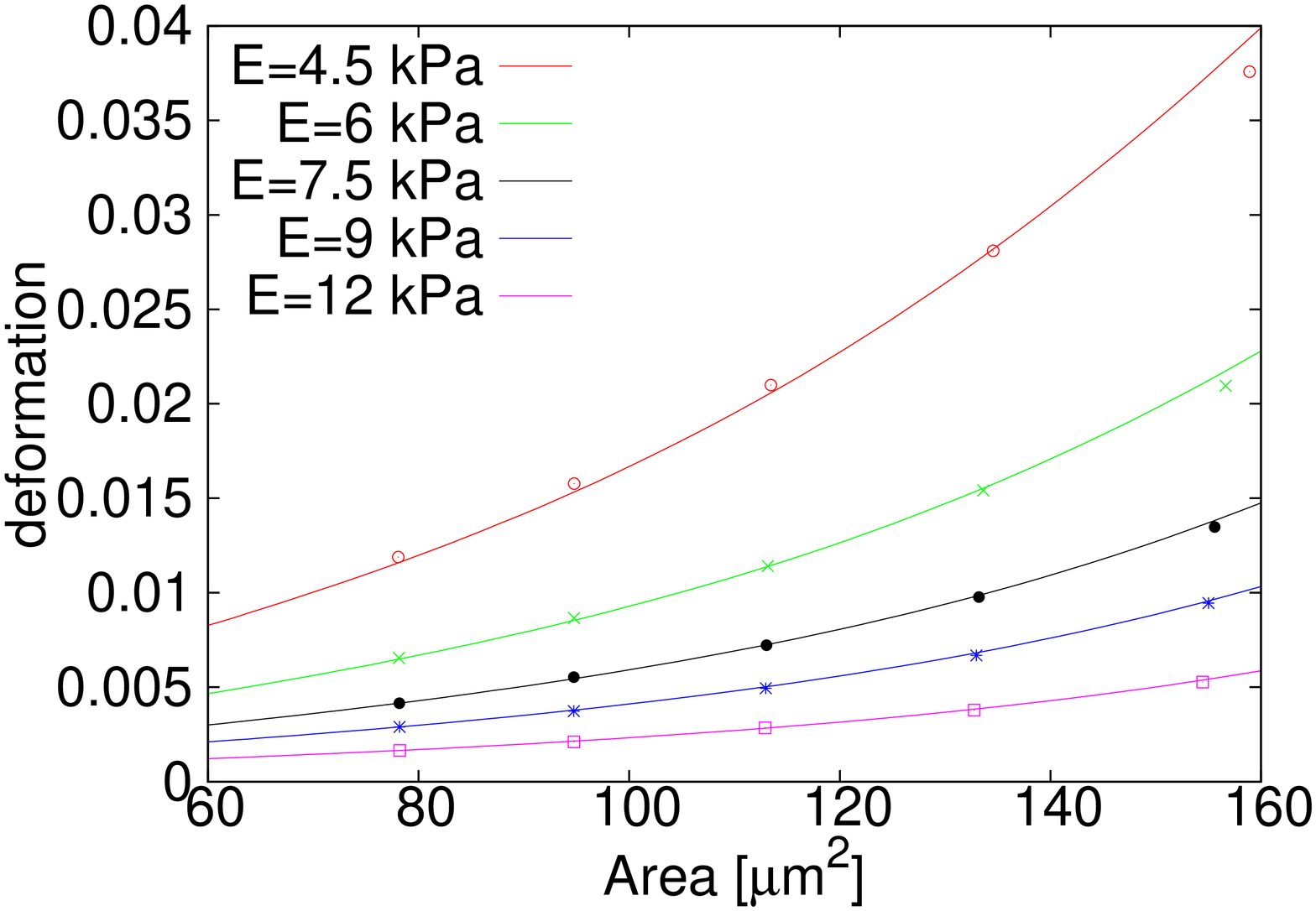} \\

(c)  &  (d)
\end{tabular}
\end{center}
\caption{Cell deformation in comparison to ALE reference values for varying parameters. 
In (a)-(c) we use the fixed physical parameters $\text{E}=6\,$kPa, $r=6$\textmu m. 
The standard model parameters are $\lambda_1 = 1\, \text{ms}$, $\epsilon = 1.25\cdot 10^{-8}$m and $m=10^{-8}$m$^3$s/kg. 
One of these parameters is varied to investigate the influence of interface thickness $\epsilon$ (a), mobility $m$ (b) and interfacial relaxation time $\lambda_1$ (c). 
(d) provides stationary deformation values at $t=2\,$ms for different cell sizes and elastic moduli. Lines depict the results of the ALE method and marker points represent those of the phase-field method. 
}

\label{fig:params}
\end{figure}

Fig.~\ref{fig:params} shows the corresponding evolution of the cell deformation. 
In \ref{fig:params}(a) we investigate the dependence of the deformation evolution on the interface width $\epsilon$. A doubling of $\epsilon$ leads to a slighty higher cell deformation. For $\epsilon= 0.0125\mu$m we find already a very good agreement to the ALE reference value, as expected from the sharp interface limit of the equations.

Next, we vary the mobility to test the influence of the intrinsic Cahn-Hilliard dynamics on the simulation results.
Therefore, we conduct a simulation with larger mobility such that the Cahn-Hilliard dynamics is doubled.
As seen in Fig. \ref{fig:params}(b), the steady-state deformation changes only slightly. This indicates that the chosen mobility is already small enough that the Cahn-Hilliard dynamics has no remarkable effect on the results. 

As a last step of the parameter study, we vary the only free parameter in the Oldroyd-B equation, $\lambda_1$.
We find that small variations of $\lambda_1$ have almost no influence on the evolution and steady state of the cell shape. To push our model to its limits, we vary $\lambda_1$ across several orders of magnitude in Fig.~\ref{fig:params}(c). We find that for very small $\lambda_1$, the cell deformation continously increases and reaches no stationary state. In this case, the very small value of $\lambda_1$ leads to a very small relaxation time in the diffuse interface region and hence to a large dissipation of elastic stress there. The drop in elastic stress lets the cell increasingly deform. On the other hand, when $\lambda_1$ is very large, a small amount of elastic stress from the diffuse interface may accumulate in the fluid, leading to increased stiffening of the fluid and a decrease in cell deformation.
We conclude that the parameter $\lambda_1$ has to be carefully chosen with a good choice being in the range of the problem's characteristic time scale.

Finally, we show that the results of the phase-field method are accurate over a range of cell sizes and elastic moduli. 
We simulated cells of five different sizes in the range $\approx 77\,\text{\textmu}\text{m}^2$ to $160\,\text{\textmu}\text{m}^2$. For each cell size, we chose five different values for the Young's  Modulus between $4.5\,$kPa and $12\,$kPa. 
As seen in Fig. \ref{fig:params}(d), the stationary deformation values of the phase-field method are in very good agreement to the ALE reference values. Only for the highest deformation values, the phase-field method underpredicts cell deformation, possibly due to small Cahn-Hilliard dynamics which becomes more prominent for larger deviations from a circular shape.

\section{Illustration of the Method's Potential} \label{sec:potential}

\begin{figure}
\begin{center}
\includegraphics[angle=-90, scale=0.18]{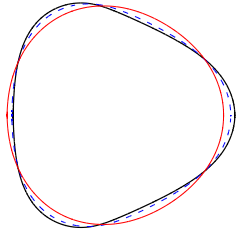}
\end{center}
\caption{Comparison of the steady state shapes for varying the surface tension ${\gamma}=5\mathrm{e}{-4}\,\text{N/m}\,$(black), 
${\gamma}=1\mathrm{e}{-3}\,\text{N/m}\,$(blue dashed) 
${\gamma}=5\mathrm{e}{-3}\,\text{N/m}$(thin red) 
}
\label{fig:surfacetension}
\end{figure}

In this section we perform further simulation studies in order to illustrate the potential of the presented phase-field FSI model.
At first, we stick to the simulations of a cell in a cylindrical channel, but we now include surface tension to demonstrate the model's capability to simulate elastic bodies with strong surface tension, as they are common in biological applications. 
Therefore, we choose three different values for the surface tension,
${\gamma}=5\mathrm{e}{-4}\,$N/m, ${\gamma}=1\mathrm{e}{-3}\,$N/m and ${\gamma}=5\mathrm{e}{-3}\,$N/m.
Further parameters are $\text{E}=3\,\text{kPa}$ and $r=6\,\text{\textmu m}$.
Fig.~\ref{fig:surfacetension} shows that the surface tension has a strong influence on the stationary cell shape, which varies from triangular (${\gamma}=5\mathrm{e}{-4}\,$N/m) to almost almost circular (${\gamma}=5\mathrm{e}{-3}\,$N/m). 
The stiff surface tension forces here are treated with a monolithic coupling of the interface advection and flow equations which relaxes any related time step restrictions \cite{ifdim}.

\begin{figure}
\begin{center}
\begin{tabular}{cc}
\includegraphics[width=0.48\textwidth]{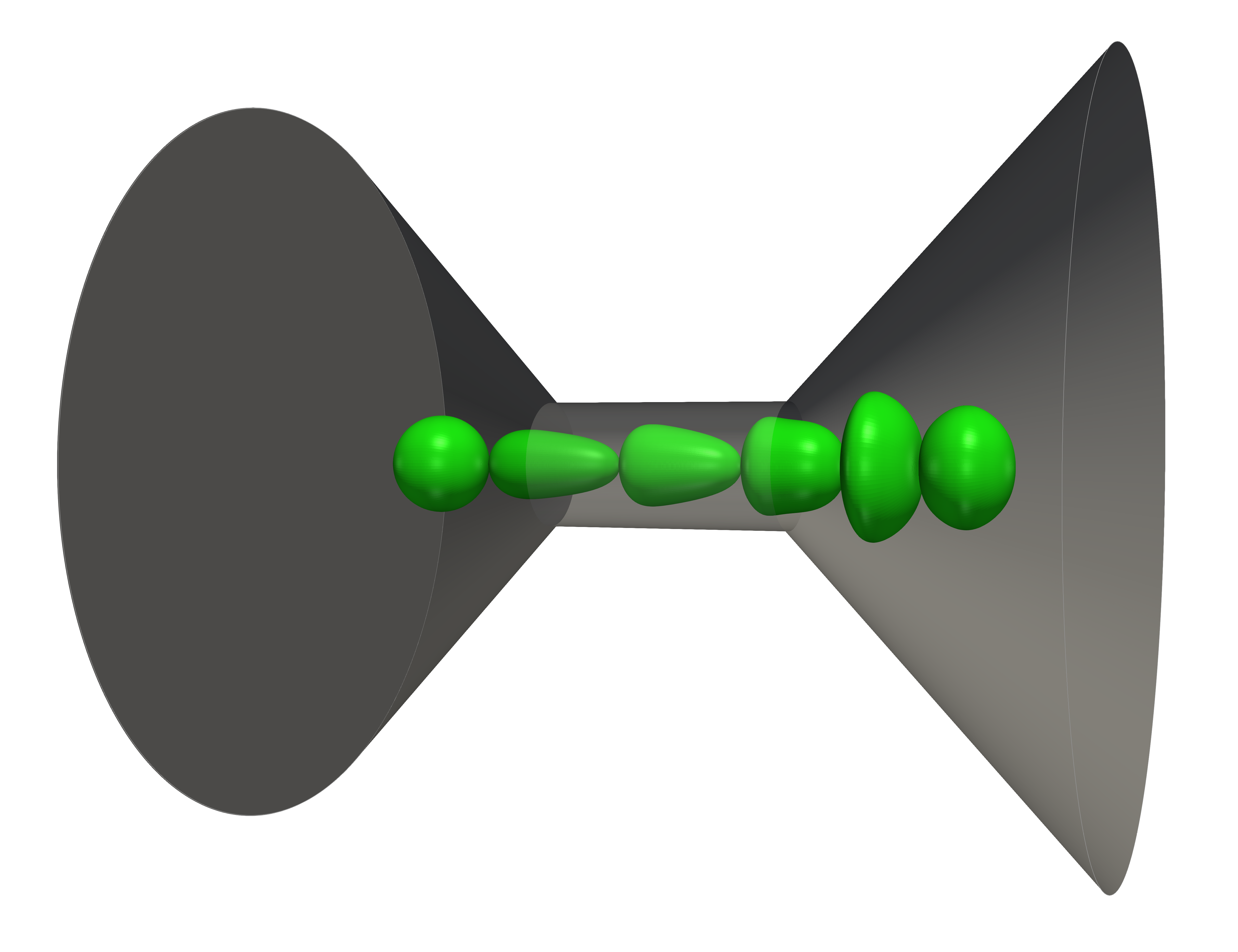} &
\includegraphics[width=0.48\textwidth]{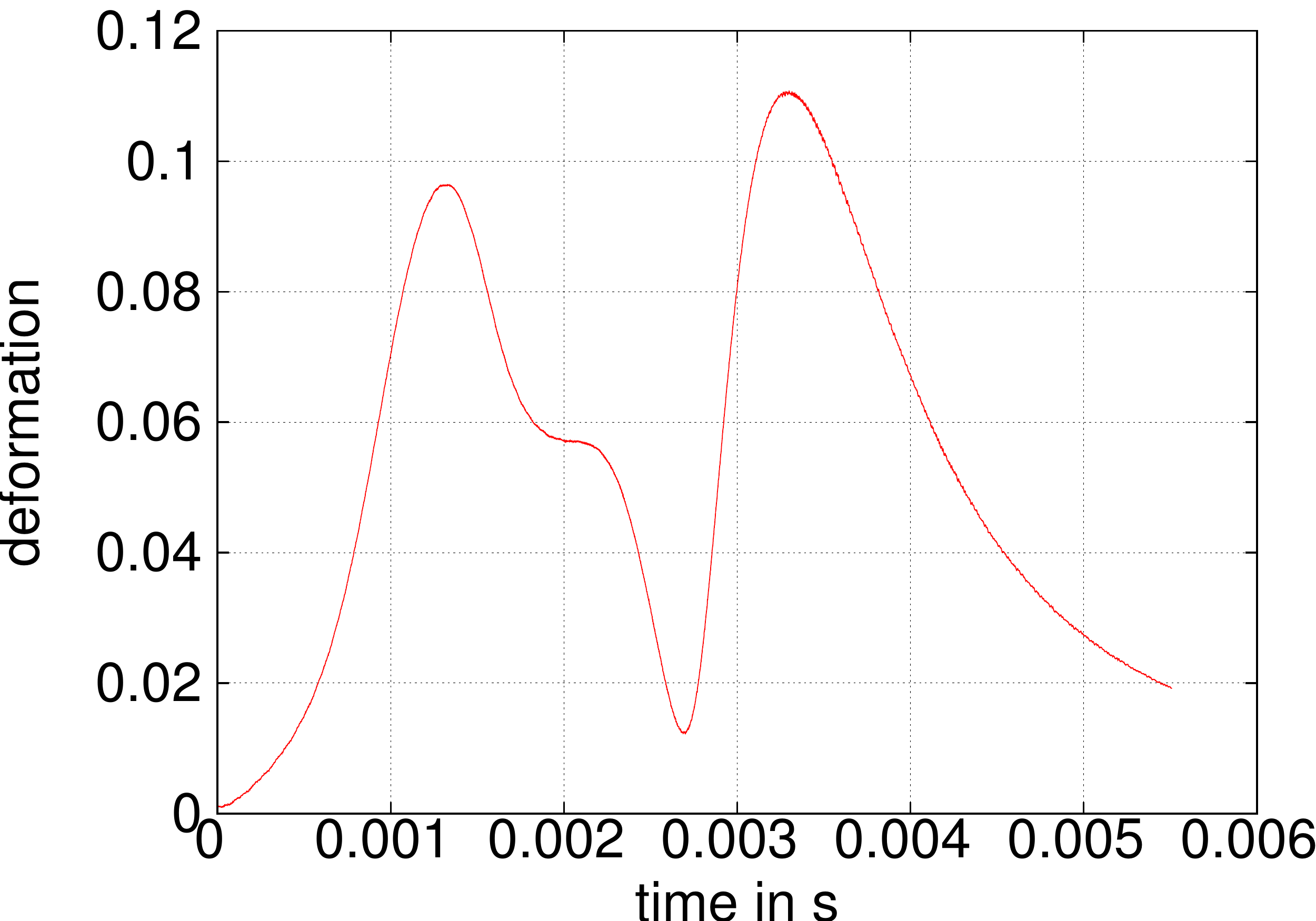}
\end{tabular}
\end{center}
\caption{Simulation of cell inflow/outflow in a modified channel geometry. The length of the narrow cylindrical channel is $40\,\text{\textmu m}$. 
\textbf{Left}: Snapshots of cell shapes at $t$=0\,ms, 1.25\,ms, 1.927\,ms, 2.595\,ms, 3.6\,ms, 5.509\,ms. 
\textbf{Right}: Cell deformation over time. 
 }
\label{fig:inflow}
\end{figure}

Next, we simulate the inflow and outflow of the cell. Such simulations are of great interest for biotechnological applications as the dynamics of cell deformation provides additional information on the cell's state. 
Therefore, we choose a realistic computation channel domain with a conical inlet/outlet of 45\textdegree \cite{Otto}. 
The chosen parameter set is: $\text{E}=1.5\,\text{kPa},\,{\gamma}=1\mathrm{e}{-3}\,$N/m$, \lambda=1\,\text{ms},\,m=10^{-8}$m$^3$s/kg$,\text{cell radius }r=8\,\text{\textmu m}\text{ and }\epsilon=0.1\mu$m. 
As for the cylindrical domain, a pressure difference is implemented between the left and the right boundary, which induces the flow. 
Fig.~\ref{fig:inflow} shows a cut through the computational domain, the initial cell position and various cell shapes during the traversal of the channel. 

The deformation curve in Fig. \ref{fig:inflow} shows a strong increase of cell deformation (elongation) during inflow, followed by a drop in deformation as soon as the cell is completely with in the cylindrical part of the channel. The elongated cell almost approaches a stationary shape around $t=2$ms, but as it is already close to the outlet, the stationary state is never reached. Instead the cell starts to become shorter and wider. This leads to a drop in the deformation, followed by a peak when the cell reaches a maximum thickness and an oblate shape as it leaves the cylindrical channel. Afterwards, the cell relaxes back to a sphere. 

Note, that such simulations are typically challenging for ALE methods as re-triangulations and interpolations are needed to reconnect the different grids while they move past each other. 
Our phase-field model needs neither re-triangulations nor interpolations to simulate this test case. 

Finally, we illustrate the capability of the phase-field method to deal with contact between an elastic material and a rigid wall. Therefore we simulate a bouncing elastic ball immersed in a fluid. 
The fluid fills a cylindrical column of height 40$\mu$m and radius 10$\mu$m. The ball of radius $r=6\mu m$ is initially placed $20\mu$m in the middle of the column. 
The parameters for this toy problem are $\nu_1=4~\mu$Pa$\cdot$s, $\nu_{-1}=10~\mu$Pa$\cdot$s, $E=500$~Pa, $\rho_1=1000$kg/m$^3$, $\rho_{-1}=100$kg/m$^3$, $\lambda_1=1$~ms, $\epsilon=0.2\mu$m, $m=2\cdot 10^{-9}$m$^3$s/kg.

A gravity force of magnitude $10^3$m/s$^2)\rho(\phi)$ is included to make the heavy ball fall down.
A no slip condition ${\bf v}=0$ is specified at the top and bottom boundary of the liquid column.
An additional no-wetting condition, $\phi=-1$, on all domain boundaries ensures that the ball is repelled from the boundaries. 
A free slip condition is imposed at the sides of the column. 

\begin{figure}
\begin{center}
\begin{tabular}{ccccc}
\includegraphics[width=0.17\textwidth]{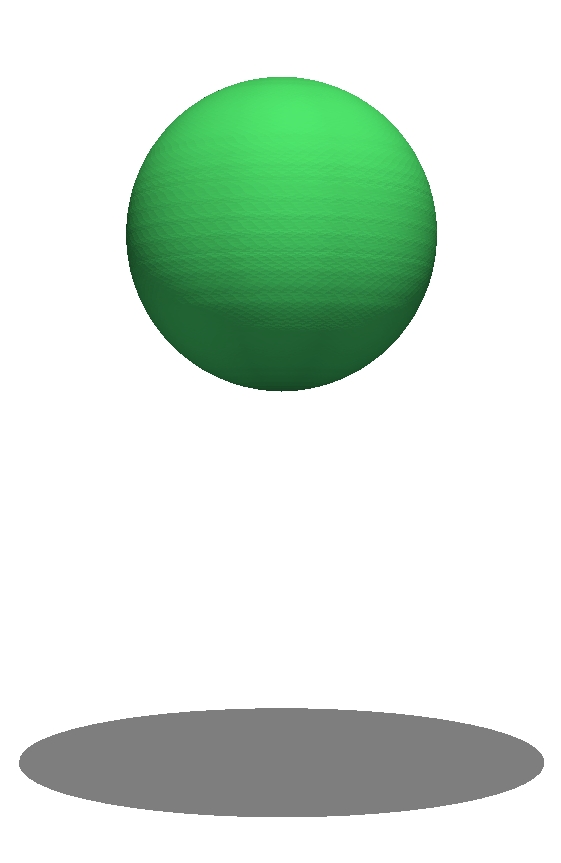} &
\includegraphics[width=0.17\textwidth]{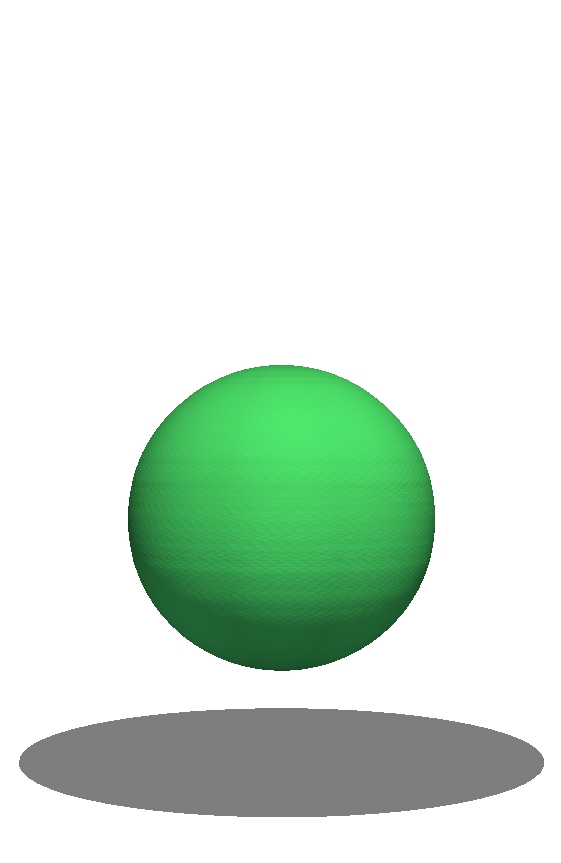} &
\includegraphics[width=0.17\textwidth]{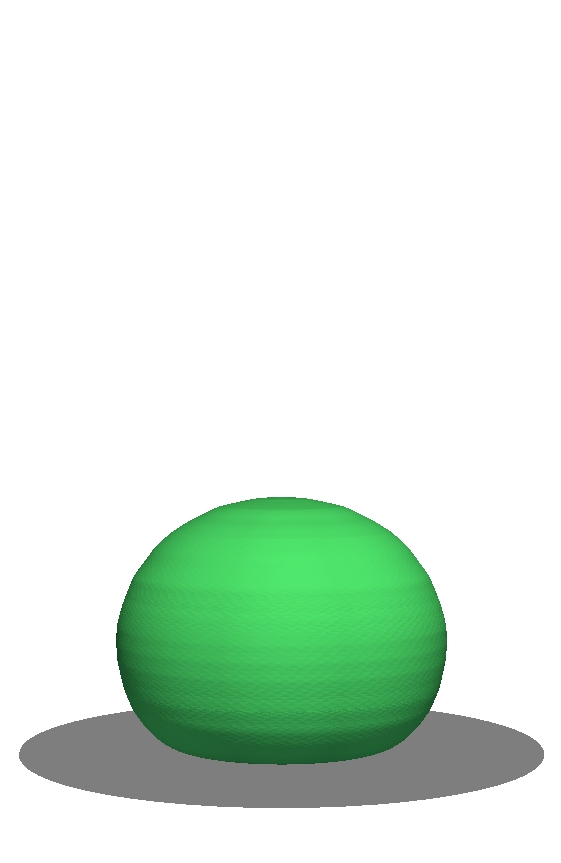} &
\includegraphics[width=0.17\textwidth]{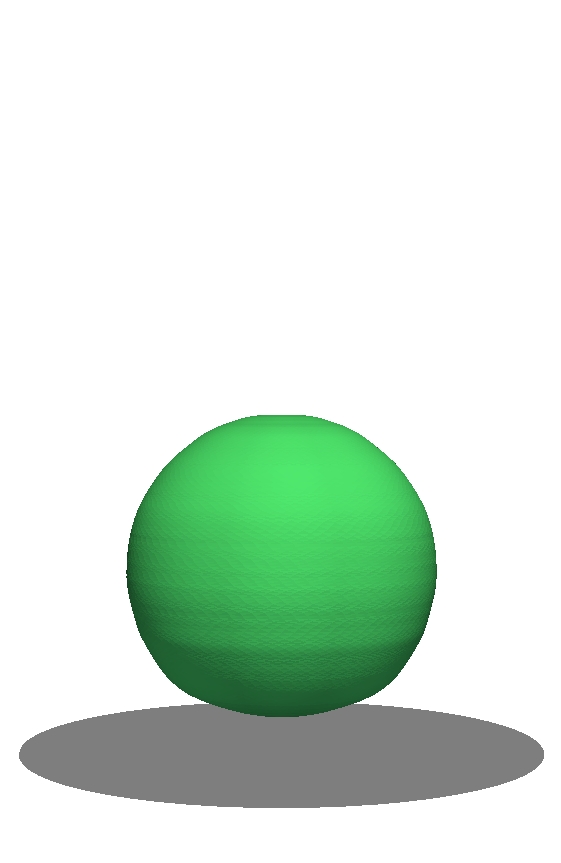} &
\includegraphics[width=0.17\textwidth]{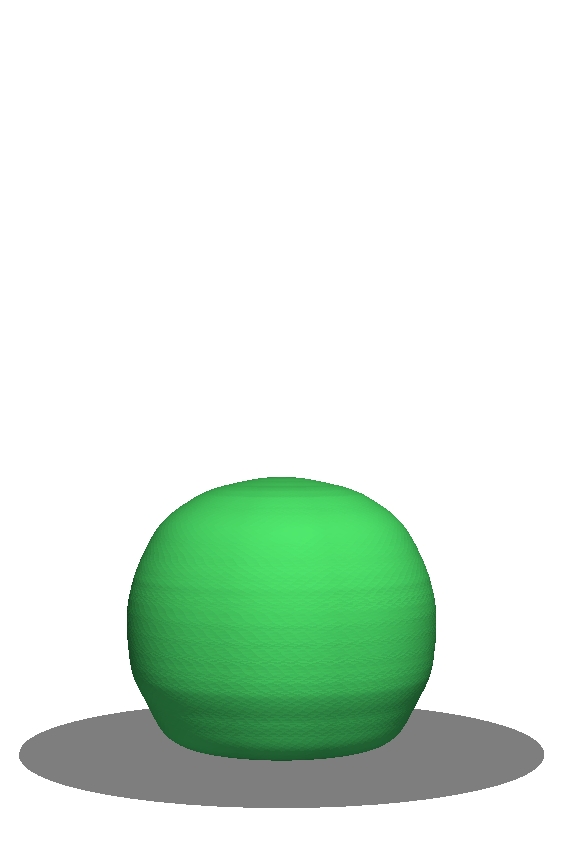} \\
$t=0$ms & $t=0.15$ms & $t=0.2$ms & $t=0.26$ms & $t=0.6$ms 
\end{tabular}
\end{center}
\caption{Simulation snapshots of an elastic ball bouncing off a rigid wall at different times.  }
\label{fig:bouncing}
\end{figure}

Snapshots of the simulation results are shown in Fig.~\ref{fig:bouncing}. 
The ball and the fluid around it are accelerated as the ball starts falling in the begin of the simulation. 
Around $t=0.18$ms the ball 'touches' the rigid wall whereupon it is compressed in the direction of motion.
After the maximum compression is reached around $t=0.2$ms the stored elastic energy is transformed into kinematic energy and the ball starts jumping upwards. This bouncing up and down is repeated several times, but quickly damped due to the viscosity of the surrounding fluid, such that the ball assumes a resting position 'lying' on the rigid wall. 

Note, that no special treatment is needed to realize the contact dynamics here. The only thing is the no-wetting condition, $\phi=-1$, which needs to be imposed at the contact boundary.

In a last step, we extend the simulation even further by adding adhesion of the ball to the substrate (wall). 
Adhesive behavior is typical for biological cells and fluids and appears as soon as the surface energy between the wall and the solid is low. 
Here, we impose equal surface energies of the fluid/wall and the solid/wall interface, which according to the 
Young-Laplace law leads to a contact angle of 90\textdegree. 
This yields a Neumann boundary condition for the phase-field, ${\bf n}\cdot\nabla\phi=0$. 
Arbitrary surface energies and contact angles can be treated as described in \cite{Aland2010,Aland_SPP_Abschluss_2017}.
 
\begin{figure}
\begin{center}
\begin{tabular}{ccccc}
\includegraphics[width=0.17\textwidth]{axi_adhesion_t0.jpg} &
\includegraphics[width=0.17\textwidth]{axi_adhesion_t150.jpg} &
\includegraphics[width=0.17\textwidth]{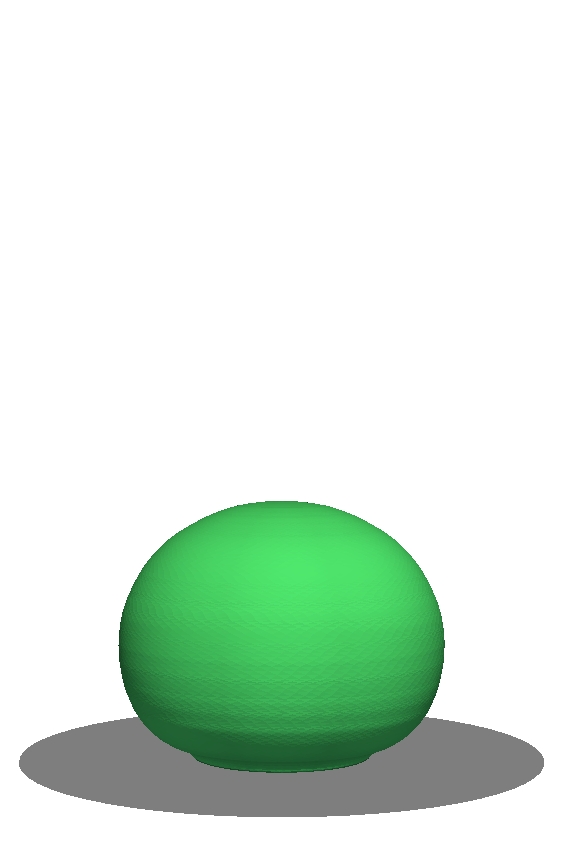} &
\includegraphics[width=0.17\textwidth]{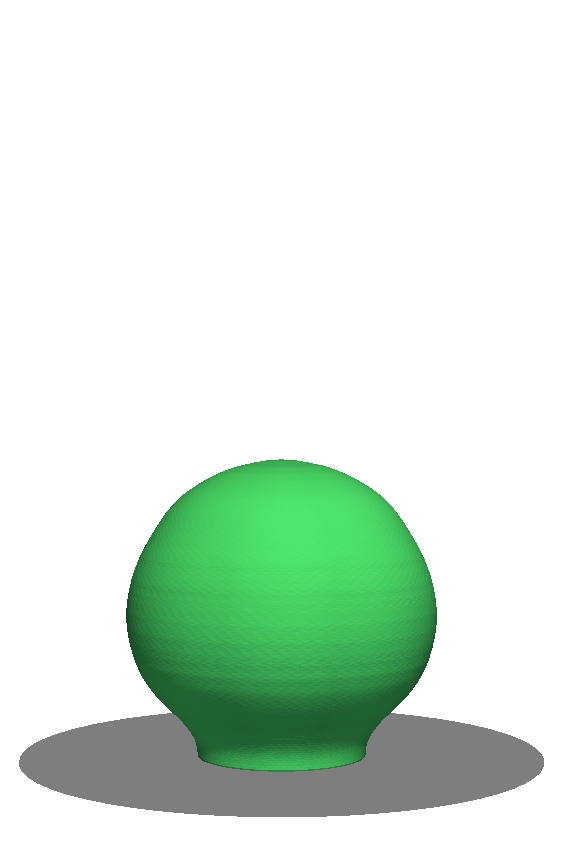} &
\includegraphics[width=0.17\textwidth]{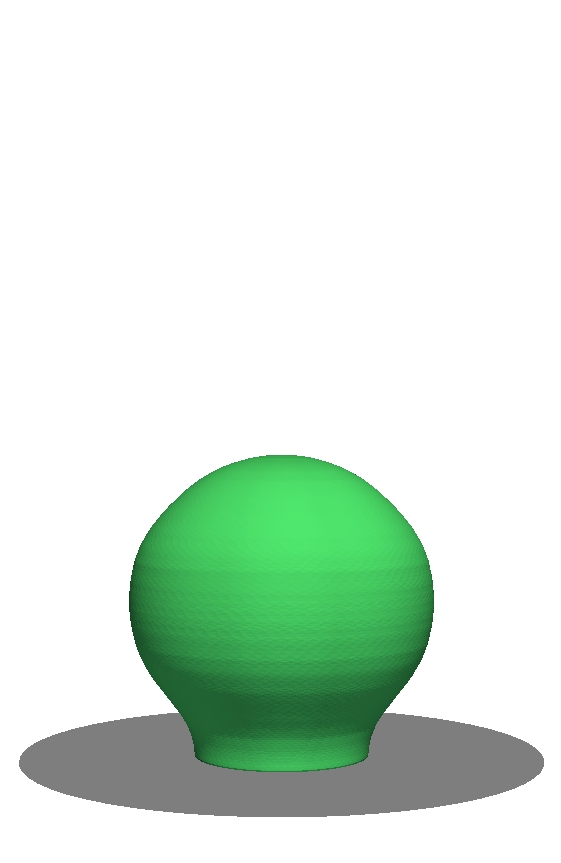}  \\
$t=0$ms & $t=0.15$ms & $t=0.2$ms & $t=0.3$ms & $t=0.4$ms 
\end{tabular}
\end{center}
\caption{Simulation snapshots of an adhesive elastic ball bumping into a rigid wall. }
\label{fig:adhesion}
\end{figure}

We repeat the simulation of the bouncing ball now with this new boundary condition to model adhesion. The parameters are as before, except for the mobility which is increased by a factor of 40 to overcome the typical stress singularity at the contact line, see \cite{Aland2010} for a discussion. 
Fig.~\ref{fig:adhesion} shows the corresponding time evolution. As soon as the ball touches the wall it starts to adhere to it with the prescribe 90\textdegree angle. 
Still, the ball is compressed shortly after contact and the elastic energy is released by lifting the ball up. But this time the ball remains bound to the wall and the oscillations are even more quickly damped. And the ball develops an almost stationary position around $t=0.4$ms. 
To our knowledge this is the first numerical method to model adhesive elastic structures of arbitrary surface energy and contact angles in flow.

\section{Conclusion} \label{sec:conclusion}

In this paper, we presented a novel phase-field model for Fluid-Structure-Interaction. 
The model is based on a monolithic Navier-Stokes equation that solves for the velocity field in both, the fluid and the elastic domain. Viscous and elastic stresses are restricted to the corresponding domains by multiplication with their characteristic functions. To obtain the elastic stress, an additional Oldroyd-B - like equation is solved including an interfacial relaxation time. 
To close the system of equations, we derived globally thermodynamically consistent forces and fluxes using energy variation arguments. Provided that suitable power series expansions exist, matched asymptotic analysis shows the convergence of the derived equations to the traditional sharp interface formulation of FSI equations.

We conducted several numerical tests to validate the applicability and accuracy of the new model.
A challenging benchmark scenario of an elastic cell traversing a fluid channel is employed and results are compared to reference values from ALE simulations\cite{Mokbel2017}. We find very good agreement for various cell sizes and elastic moduli. In particular, we show that the interface thickness $\epsilon$ and the mobility $\gamma$ are small enough to influence the results only marginally. 
Results are also shown to be robust with respect to the introduced interfacial relaxation time.

Finally, we highlight some distinct advantages of the new model as compared to traditional ALE approaches for FSI. 
We demonstrate the movement of a solid object through a fluidic channel without grid re-triangulations. 
We include strong surface energy, i.e. surface tension forces, into the model, whose stable discretization is one of the advantages of phase-field models. 
At last, we show how easy it is to include contact dynamics into the model. 
We demonstrate this by simulating a ball bouncing off a wall.
We conclude with the simulation of adhesion of an elastic ball to a rigid wall, a scenario which, to our knowledge, cannot be simulated with any other FSI model so far. 

While we have restricted simulations to fluid-structure interaction, the model is capable to simulate any combination of viscous fluids, visco-elastic fluids and elastic solids. 
We therefore believe that the proposed phase-field model is well suited to tackle a range of complicated multi-physics problems, in particular from biology, in the future.

\textbf{Acknowledgements}
SA acknowledges support from the German Science Foundation (grant AL 1705/3) and support from the Saxon Ministry for Science and Art (SMWK MatEnUm-2).

\section{Appendix}
\label{appendix}

\subsection{Energy Time Derivative} \label{sec: energy time derivative}
In the following we present the complete computation of the time evolution of the energy given in Section \ref{sec:EnergyDissipation}. As discussed in \cite{Boyaval}, we assume $\sigma\in\mathbb{R}^{n\times n}$ to be a symmetric positive definite matrix and thus calculate the trace of the matrix logarithm $\ln\sigma$ by
\begin{align}
\text{tr}\left(\ln\sigma\right)=\sum_{i=1}^n\ln\lambda_i~
\end{align}
with the eigenvalues $\lambda_i$. 
According to Eq. \eqref{energy}, we consider three parts of the total energy:
\begin{align}
E=\int_{\Omega}\underbrace{\frac{\rho(\phi)}{2}\left| \mathbf{v}\right|^2}_\text{kinetic energy $E_{\text{kin}}$}+\underbrace{\frac{\mu(\phi)}{2}\text{tr}\left(\mathbf{\sigma}-\ln\mathbf{\sigma}-\mathbb{I}\right)}_\text{elastic energy $E_{\text{el}}$}+\underbrace{\tilde{\gamma}\left(\frac{\epsilon}{2}\left|\nabla\phi\right|^2+\frac{1}{\epsilon}W(\phi)\right)}_\text{Cahn-Hilliard energy $E_{\text{CH}}$}~\text{d}x~.
\end{align}  
The following identities will be useful to compute the time derivative of the three energies:
\begin{align}
\nabla{\bf v}:\left(\nabla{\bf v}+\nabla{\bf v}^T\right)&=\frac{1}{2}\left|\nabla{\bf v}+\nabla{\bf v}^T\right|^2 \label{id1} \\
\partial^{\bullet}\rho(\phi)&=\rho '(\phi)\partial^{\bullet}\phi \label{id3} \\
\partial^{\bullet}\left(\frac{\rho(\phi)}{2}\left|{\bf v}\right|^2\right)&=\frac{1}{2}\partial^{\bullet}\left(\rho(\phi)\right)\left|{\bf v}\right|^2+\partial^{\bullet}\left({\bf v}\right)\cdot\rho(\phi){\bf v} \notag \\
&= \partial^{\bullet}\left(\rho(\phi){\bf v}\right)\cdot{\bf v}-\partial^{\bullet}\rho(\phi)\frac{\left|{\bf v}\right|^2}{2}\label{id4}
\end{align}

In the following computation of the energy time derivatives, we will neglect boundary integrals that arise from integration by parts, as we assume appropriate boundary conditions. We compute,

{\footnotesize
\begin{align*}
\frac{d}{dt} E_{\text{kin}}&{\underset{\eqref{id4}}{=}}\int_{\Omega}\partial^{\bullet}(\rho(\phi){\bf v})
{\bf v}-\partial^{\bullet}\frac{\rho(\phi)}{2}\left|{\bf v}\right|^2~\text{d}x \\
&{\underset{\eqref{navier stokes}}{=}}\int_{\Omega}\left(\textbf{F}+\nabla\cdot\left(\nu(\phi)\left(\nabla{\bf v}+
\nabla{\bf v}^{\text{T}}\right)\right)+\nabla p+\nabla\cdot\left(\mu(\phi)\sigma\right)\right)\cdot{\bf v}-\partial^{\bullet}\frac{\rho(\phi)}{2}\left|{\bf v}\right|^2~\text{d}x \\
&= \int_{\Omega}\textbf{F}\cdot{\bf v}-\nu(\phi)\left(\nabla{\bf v}+
\nabla{\bf v}^{\text{T}}\right):\nabla{\bf v}-p\nabla\cdot{\bf v}-\mu(\phi)\sigma :\nabla{\bf v}-\partial^{\bullet}\frac{\rho(\phi)}{2}\left|{\bf v}\right|^2~\text{d}x \\
&{\underset{\eqref{transport}-\eqref{id3}}{=}}\int_{\Omega}-\frac{\nu(\phi)}{2}\left|\nabla{\bf v}+
\nabla{\bf v}^{\text{T}}\right|^2-\mu(\phi)\sigma :\nabla{\bf v}+{\bf v}\cdot \textbf{F}+\frac{\rho'(\phi) }{2}\left|{\bf v}\right|^2\nabla\cdot \textbf{J}~\text{d}x,
\\
& \notag
\\
\frac{d}{dt} E_{\text{CH}} &= \int_{\Omega}\tilde{\gamma}\left(-\epsilon\Delta\phi+\frac{1}{\epsilon}W'(\phi)\right)\partial^{\bullet}\phi+{\bf v}
\cdot\left(\nabla\cdot\left(\epsilon\tilde{\gamma}\nabla\phi\otimes\nabla\phi\right)\right)~\text{d}x \\
&{\underset{\eqref{transport}}{=}}\int_{\Omega}-\frac{\delta E_{\text{CH}}}{\delta\phi}\nabla\cdot \textbf{J}+{\bf v}\cdot(\nabla\cdot\left(\epsilon\tilde{\gamma}\nabla\phi\otimes\nabla\phi\right))~\text{d}x ~,
\end{align*}
}

where we defined $\frac{\delta E_{\text{CH}}}{\delta\phi}:=\tilde{\gamma}\left(-\epsilon\Delta\phi+\frac{1}{\epsilon}W'(\phi)\right)$. 

{
\small
\begin{align*}
\frac{d}{dt} E_{\text{el}}&=\int_{\Omega}\partial^{\bullet}\left(\frac{\mu(\phi)}{2}\right)\text{tr}\left(\mathbf{\sigma}-\ln\mathbf{\sigma}
-\mathbb{I}\right)+\frac{\mu(\phi)}{2}\partial^{\bullet}~\text{tr}\left(\mathbf{\sigma}-\ln\mathbf{\sigma}
-\mathbb{I}\right)~\text{d}x \\
&\hspace{-0.35cm}{\underset{\eqref{transport},\eqref{id3}}{=}}\int_{\Omega}-\frac{\mu'(\phi)}{2}\nabla\cdot \textbf{J}~\text{tr}\left(\mathbf{\sigma}-\ln\mathbf{\sigma}
-\mathbb{I}\right)+\frac{\mu(\phi)}{2}\text{tr}~\left(\partial^{\bullet}\left(\sigma-
\ln\sigma\right)\right) ~\text{d}x\\
&=\int_{\Omega}-\frac{\mu'(\phi)}{2}\nabla\cdot \textbf{J}~\text{tr}\left(\mathbf{\sigma}-\ln\mathbf{\sigma}
-\mathbb{I}\right)  ~\text{d}x +\int_{\{\lambda=0\}}\frac{\mu(\phi)}{2}\text{tr}~\left((\mathbb{I}-\sigma^{-1})\partial^{\bullet}\sigma\right) ~\text{d}x \\
&~~~~~~+\int_{\Omega\backslash\{\lambda=0\}}\frac{\mu(\phi)}{2}\text{tr}~\left((\mathbb{I}-\sigma^{-1})\partial^{\bullet}\sigma\right) ~\text{d}x
\\
&{\underset{\eqref{B evolution combined}}{=}}\int_{\Omega}-\frac{\mu'(\phi)}{2}\nabla\cdot \textbf{J}~\text{tr}\left(\mathbf{\sigma}-\ln\mathbf{\sigma}
-\mathbb{I}\right)  ~\text{d}x \\ 
&~~~~~~+\int_{\Omega\backslash\{\lambda=0\}}\frac{\mu(\phi)}{2}~\text{tr}~\left(\left(\mathbb{I}-\sigma^{-1}\right)
\left(\nabla{\bf v}^T\cdot\sigma+\sigma\cdot
\nabla{\bf v}-\frac{\alpha(\phi)}{\lambda(\phi)}\left(\sigma-\mathbb{I}\right)\right)\right)~\text{d}x.\notag
\end{align*}
}

\noindent
Note, that Eq.~\eqref{B evolution combined} yields the boundedness of the last integrand  in the set ${\Omega\backslash\{\lambda=0\}}$ given the solution is sufficiently smooth. 
Consequently, we obtain the variation of $E$ as
\begin{align}
\text{d}_tE&=\int_{\Omega}-\frac{\nu(\phi)}{2}\left|\nabla{\bf v}+\nabla{\bf v}^T\right|^2-\mu(\phi)\sigma:\nabla{\bf v}+{\bf v}\cdot\textbf{F}+\frac{\rho'(\phi)}{2}\left|{\bf v}\right|^2\nabla\cdot \textbf{J} \label{dtE1} \\
&~~~~~
-\frac{\mu'(\phi)}{2}\nabla\cdot \textbf{J}~\text{tr}\left(\sigma-\ln\sigma-\mathbb{I}\right)-\nabla\cdot \textbf{J}\frac{\delta E_{\text{CH}}}{\delta\phi}+{\bf v}\cdot(\nabla\cdot\left(\epsilon\tilde{\gamma}\nabla\phi\otimes\nabla\phi\right)) ~\text{d}x \notag \\
&~~~ +\int_{\Omega\backslash\{\lambda=0\}} \frac{\mu(\phi)}{2}~\text{tr}\left(\left(\mathbb{I}-\sigma^{-1}\right)
\left(\nabla{\bf v}^T\sigma+\sigma\nabla
{\bf v}-\frac{\alpha(\phi)}{\lambda(\phi)}\left(\sigma-\mathbb{I}\right)\right)\right)~\text{d}x~. \notag
\end{align}

In order to reformulate the trace term in the last line, we use the symmetry of $\sigma$ and the following properties:
\begin{align}
\text{tr}\left(ABA^{-1}\right)&=\text{tr}\left(B\right) ~~\text{for a regular matrix A} \label{tr1} \\
\text{tr}\left(AB\right)&=A:B^T=A^T:B \label{tr2}\\
\text{tr}\left(\nabla{\bf v}\right)&=\nabla\cdot{\bf v}=0 \label{tr3}
\end{align} 
and thus get for $\lambda\neq 0$
\begin{align}
\text{tr}&\left(\left(\mathbb{I}-\sigma^{-1}\right)
\left(\nabla{\bf v}^T\sigma+\sigma\nabla
{\bf v}-\frac{\alpha(\phi)}{\lambda(\phi)}\left(\sigma-\mathbb{I}\right)\right)\right) \notag \\
&{\underset{\eqref{tr1},\eqref{tr2}}{=}}2\sigma:\nabla{\bf v}-2\text{tr}\left(\nabla{\bf v}\right)-\frac{\alpha(\phi)}{\lambda(\phi)}\text{tr}\left(\sigma+\sigma^{-1}-2\mathbb{I}\right) \nonumber \\
&{\underset{\eqref{tr3}}{=}}2\sigma:\nabla{\bf v}-\frac{\alpha(\phi)}{\lambda(\phi)}\text{tr}\left(\sigma+\sigma^{-1}-2\mathbb{I}\right)~. \label{eq 7.11}
\end{align}
Note, that also the last term of Eq.~\eqref{eq 7.11} is bounded, which follows from Eq.~\eqref{B evolution combined} and $|\text{tr} (\sigma+\sigma^{-1}-2I)|\leq C\|\sigma-I\|$ in a neighborhood of the identity matrix.
We obtain a simplified version of Eq.~\eqref{dtE1}:
\begin{align*} 
\text{d}_tE&=\int_{\Omega}-\frac{\nu(\phi)}{2}\left|\nabla{\bf v}+\nabla{\bf v}^T\right|^2+{\bf v}\cdot\textbf{F}+\frac{\rho'(\phi)}{2}\left|{\bf v}\right|^2\nabla\cdot \textbf{J} -\nabla\cdot \textbf{J}\frac{\delta E_{\text{CH}}}{\delta\phi}\\ 
&~~~~~ -\frac{\mu'(\phi)}{2}\nabla\cdot \textbf{J}~\text{tr}\left(\sigma-\ln\sigma-\mathbb{I}\right) +{\bf v}\cdot(\nabla\cdot\left(\epsilon\tilde{\gamma}\nabla\phi\otimes\nabla\phi\right))~\text{d}x \\
&- \int_{\Omega\backslash\{\lambda=0\}} \frac{\mu(\phi)\alpha(\phi)}{2\lambda(\phi)}\text{tr}\left(\sigma+\sigma^{-1}-2\mathbb{I}\right) ~\text{d}x
\end{align*}
Furthermore we reformulate the density term in \eqref{dtE1} applying integration by parts twice and using that $\rho '(\phi) $ is constant:
\begin{align*}
\int_{\Omega}\frac{\rho'(\phi)}{2}\left|{\bf v}\right|^2\nabla\cdot \textbf{J}=\int_{\Omega}-\rho'(\phi){\bf v}\cdot\left(\nabla{\bf v}\cdot \textbf{J}\right)=\int_{\Omega}{\bf v}\cdot(\nabla\cdot\left(\rho'(\phi){\bf v}\otimes \textbf{J}\right))~\text{d}x~,
\end{align*}
which yields the energy time derivative given in Eq.~\eqref{dt energy 1}.



\subsection{Time-discrete axisymmetric equations}
\label{sec:timediscrete}

\begin{figure}
\begin{center}
\begin{tabular}{ccc}
\includegraphics[width=0.7\textwidth]{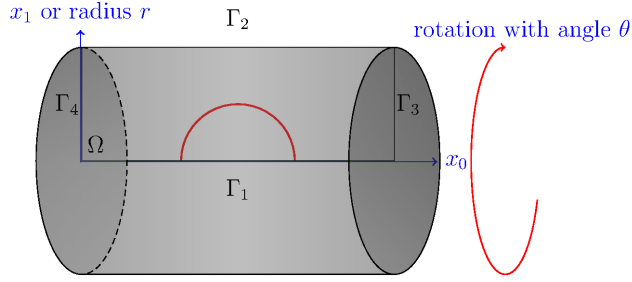}
\end{tabular}
\end{center}
\caption{Illustration of the axisymmetric computational domain. The boundary line segments $\Gamma_1$ to $\Gamma_4$ circumscribe the rectangular 2D domain $\Omega_{2\text{D}}$. The use of axisymmetric terms in the governing system of equations, allows to simulate the flow of an initially spherical object through a cylinder.}
\label{fig:setup}
\end{figure}

We consider axisymmetric flow and geometries which allows to rewrite the equations in a 2D manner using cylindrical coordinates.
Thereby the 2D meridian domain $\Omega_{2\text{D}} = \{(x_0,r) |~ 0\!\!\leq\!\! x_0 \!\!\leq\!\! a, 0\!\!\leq\!\! r \!\!\leq\!\! b\}$ 
represents the 3D domain $\Omega = \{(x_0,x_1,x_2) |~ x_1\!=\!r\cos(\theta), x_2\!=\!r\sin(\theta), (x_0,r)\in \Omega_{2\text{D}}, \theta\in[0,2\pi)\}$, see Fig.~\ref{fig:setup} for an illustration.
In the following, all fields are defined on $\Omega_{2\text{D}}$, the velocity field on this domain is defined to consist of only axial and radial component ${\bf v}=(v_0,v_r)$.
The gradient, divergence and Laplace operator in the cylindrical coordinate system are defined by
\begin{align*}
\nabla=(\partial_{x_0}, \partial_r), \qquad
\tilde{\nabla}\cdot &= \left(\partial_{x_0}, \frac{1}{r} + \partial_r\right), \qquad
\tilde{\Delta} = \tilde{\nabla}\cdot\nabla  = \partial_{x_0 x_0} + \partial_{rr} + \frac{1}{r}\partial_r. 
\end{align*}
As derivatives in azimuthal ($\theta$-) direction vanish, the strain tensor assumes the form 
\begin{align*}
{\sigma}:=\begin{pmatrix} \sigma_{00} & \sigma_{01} & 0 \\ \sigma_{10} & \sigma_{11}  & 0 \\ 0 & 0 & \sigma_{\theta\theta} \end{pmatrix}.
\end{align*}
For a shorter notation we introduce the matrices 
\begin{align*}
{\sigma}_{2\text{D}}=\begin{pmatrix} \sigma_{00} & \sigma_{01} \\ \sigma_{10} & \sigma_{11} \end{pmatrix}, \qquad 
{\mathbb{I}}_{2\text{D}}=\begin{pmatrix} 1 & 0 \\ 0 & 1 \end{pmatrix}
\end{align*}

We assume the density $\rho$ to be constant. Hence, we may neglect the corresponding term in Eq. \eqref{F}. 
The lower, upper, right, and left boundaries are denoted by $\Gamma_1$, $\Gamma_2$, $\Gamma_3$ and $\Gamma_4$, respectively. 
In the comparison study with the ALE simulations, the computational domain is moved along with the cell velocity, i.e.,
the spatially averaged velocity of the cell, named ${\bf v}_b $, is subtracted from the velocity in the advection terms in Eqs. \eqref{transport}, \eqref{navier stokes} and \eqref{B evolution combined}. 
This modification helps to reduce the amount of remeshing and also leads to a consistent comparison with the ALE model which also applied such a co-moving grid. 

For the time discretization we chose an equidistant time partitioning with time step size $\tau$. Thereby we consider the discrete time derivative $d_tf^m:=\frac{f^m-f^{m-1}}{\tau}$ for some scalar variable $f$ and time step index $m$.
For simplification, we denote the viscous stress tensor $\textbf{D}({\bf v})=\frac{1}{2}\left(\nabla {\bf v}+\nabla {\bf v}^T\right)$ in the following. 
At each timestep we solve the following systems in $\Omega_{2\text{D}}$:

\begin{enumerate}
\item The Navier-Stokes system
\begin{align}
\rho &\left(d_t{\bf v}^m+\left( {\bf v}^{m-1}-{\bf v}_b\right)\cdot\nabla {\bf v}^m\right)-\nabla p^m-2\tilde{\nabla}\cdot\left(\nu^{m-1}\textbf{D}({\bf v}^{m})\right)\notag\\
&=\frac{\nu^{m-1}}{r} \begin{pmatrix} 0 \\ - \frac{2}{r}v_1^m \end{pmatrix}
+\tilde{\nabla}\cdot\left(\mu^{m-1}\left(\sigma_{2\text{D}}^{m-1}-\mathbb{I}_{2\text{D}}\right)\right)\notag
-(\sigma_{\theta\theta}^{m-1}-1)\frac{\mu^{m-1}}{r}\begin{pmatrix} 0 \\ 1 \end{pmatrix} \\
& ~~~ - \epsilon\tilde{\gamma}\tilde{\nabla}\cdot\left(\nabla\phi^{m-1}\otimes\nabla\phi^{m-1}\right) \notag \\
&\tilde{\nabla}\cdot {\bf v}^m=0
\end{align}
where $\nu^m=\nu(\phi^m),~\mu^m=\mu(\phi^m)$.
We apply the following boundary conditions for the velocity:
\begin{align}
\label{eq:noslip}
{\bf v}&=0~~~~~~~~~~\text{ on }\Gamma_2~, \\
\label{eq:noradial}
v_1&=0~~~~~~~~~~\text{ on }\Gamma_1~.
\end{align}
Equation \eqref{eq:noslip} corresponds to a no-slip condition at the channel wall and (\ref{eq:noradial}) avoids a radial flow at the symmetry axis. 
In case of channel flow, we set periodic boundary conditions for ${\bf v}$ on $\Gamma_4$ and $\Gamma_3$ and 
\begin{align*}
p&=0&\text{ on }\Gamma_3~, \\
p&=p_0&\text{ on }\Gamma_4~, 
\end{align*}
where $p_0>0$ imposes the desired pressure difference between $\Gamma_4$ and $\Gamma_3$ driving the flow through the channel. 

\item The Cahn-Hilliard system
\begin{align}
d_t\phi^m+\left( {\bf v}^m-{\bf v}_b\right)\cdot\nabla\phi^{m-1}-m(\phi)\tilde{\Delta} c^m &=0\\
c^m+\epsilon\tilde{\Delta}\phi^m-\frac{1}{\epsilon}W'(\phi ^m)&=0
\end{align}
To avoid the nonlinear terms, we choose a Taylor expansion of linear order for $W'(\phi^m)$:
\begin{align}
W'\left(\phi^m\right) &= \left(\phi^m\right)^3-\phi^m \notag \\
&\approx 3\left(\phi^{m-1}\right)^2\phi^m-2\left(\phi^{m-1}\right)^3-\phi^m.
\end{align}
As for the velocity, periodic boundary conditions for $\phi$ and $c$ are given on $\Gamma_3$ and $\Gamma_4$.
No flux conditions are used on the other boundaries. 

\item The Oldroyd-B system
\begin{align}
\lambda(\phi^m)&\left(d_t\sigma_{2\text{D}}^m+\left({\bf v}^m-{\bf v}_b\right)\nabla\sigma_{2\text{D}}^m-\nabla{\bf v}^{m}\cdot\sigma_{2\text{D}}^{m-1}-\sigma_{2\text{D}}^{m-1}\cdot\left(\nabla{\bf v}^m\right)^T\right)\notag \\
&=D\left(\Delta\sigma_{2\text{D}}^m-\Delta\sigma_{2\text{D}}^{m-1}\right) -\alpha(\phi^m)\left(\sigma_{2\text{D}}^m-\mathbb{I}\right)\label{eq:ob-1}\\
\lambda(\phi^m)&\left(d_t\sigma^m_{\theta\theta}+\left({\bf v}^m-{\bf v}_b\right)\nabla\sigma_{\theta\theta}^m-2\frac{v_1^m}{r}\sigma_{\theta\theta}^{m-1}\right)\notag\\
&=D\left(\Delta\sigma^m_{\theta\theta}-\Delta\sigma^{m-1}_{\theta\theta}\right) -\alpha(\phi^m)\left(\sigma^m_{\theta\theta}-\mathbb{I}\right).\label{eq:ob-2}
\end{align}
 Note that \eqref{eq:ob-1}-\eqref{eq:ob-2} express 5 equations for the 5 unknowns of the elastic stress tensor. To ensure numerical stability we have added a small artificial diffusion term with $D=2\cdot 10^{-10}$m$^2$, whereupon natural no-flux boundary conditions emerge.
\end{enumerate}

\noindent
{\bf\large References}
\bibliographystyle{abbrv} 
\bibliography{Bibliographie20171221} 

\end{document}